\documentclass{article}
\usepackage{arxiv}
\usepackage[T1]{fontenc}    

\usepackage[utf8]{inputenc}
\usepackage{graphicx}
\usepackage{xcolor}
\usepackage{microtype}
\usepackage{listings}

\definecolor{codegreen}{rgb}{0,0.6,0}
\definecolor{codegray}{rgb}{0.5,0.5,0.5}
\definecolor{codepurple}{rgb}{0.58,0,0.82}
\definecolor{backcolour}{rgb}{0.95,0.95,0.92}

\usepackage{bbm}
\usepackage{amsmath,amssymb,amsfonts,amsthm}
\usepackage{siunitx}
\usepackage{booktabs}
\usepackage{multirow}

\usepackage{hyperref}

\theoremstyle{remark} 
\newtheorem*{remark}{Remark}

\renewcommand{\vec}[1]{\boldsymbol{#1}}

\newcommand{\dd}{\mathrm{d}}
\newcommand{\pfrac}[2]{\frac{\partial #1}{\partial #2}}

\title{A highly efficient computational approach for fast scan-resolved simulations of metal additive manufacturing processes on the scale of real parts}

\newcommand{\shorttitle}{A highly efficient computational approach for fast scan-resolved simulations of metal AM processes on the part-scale}

\author{Sebastian D.~Proell\thanks{corresponding author}\\
	Institute for Computational Mechanics\\
	Technical University of Munich\\
	85748 Garching b. München\\
	\texttt{sebastian.proell@tum.de} \\
	\And
	Peter Munch\\
	Institute of Mathematics\\
	University of Augsburg\\
	86159 Augsburg\\
	\AND
	Martin Kronbichler\\
	Institute of Mathematics\\
	University of Augsburg\\
	86159 Augsburg\\
	\AND
	Wolfgang A.~Wall\\
	Institute for Computational Mechanics\\
	Technical University of Munich\\
	85748 Garching b. München\\
	\And
	Christoph Meier\\
	Institute for Computational Mechanics\\
	Technical University of Munich\\
	85748 Garching b. München
}

\begin{document}
	\maketitle
	\begin{abstract}
		This article proposes a novel high-performance computing approach for the prediction of the temperature field in powder bed fusion (PBF) additive manufacturing (AM) processes. In contrast to many existing approaches to part-scale simulations, the underlying computational model consistently resolves physical scan tracks without additional heat source scaling, agglomeration strategies or any other heuristic modeling assumptions. A growing, adaptively refined mesh accurately captures all details of the laser beam motion.
		Critically, the fine spatial resolution required for resolved scan tracks in combination with the high scan velocities underlying these processes mandates the use of comparatively small time steps to resolve the underlying physics. Explicit time integration schemes are well-suited for this setting, while unconditionally stable implicit time integration schemes are employed for the interlayer cool down phase governed by significantly larger time scales. These two schemes are combined and implemented in an efficient fast operator evaluation framework providing significant performance gains and optimization opportunities. The capabilities of the novel framework are demonstrated through realistic AM examples on the centimeter scale including the first scan-resolved simulation of the entire NIST AM Benchmark cantilever specimen, with a computation time of less than one day. Apart from physical insights gained through these simulation examples, also numerical aspects are thoroughly studied on basis of weak and strong parallel scaling tests. As potential applications, the proposed thermal PBF simulation approach can serve as a basis for microstructure and thermo-mechanical predictions on the part-scale, but also to assess the influence of scan pattern and part geometry on melt pool shape and temperature, which are important indicators for well-known process instabilities.
	\end{abstract}
	
	
	\section{Introduction}
	Metal additive manufacturing (AM) offers a variety of advantages over conventional manufacturing techniques \cite{gibson2021additive,Herzog2016}. This contribution focuses on powder bed fusion AM (PBFAM) where the desired part geometry is molten into a powder bed by means of a laser (or electron) beam. However, the approach presented in this article is also transferable to other processes such as directed energy deposition (DED).
	
	One of the most commonly cited advantages of AM is the ability to produce complex geometries in a near net shape manner. As exciting as this promise may be for the industry as a whole, it also poses new challenges for part design: due to the high geometrical complexity a part may not be manufacturable with the desired quality or adequate process parameters are hard to find. Various defects such as porosity, dimensional warping and delamination are known in the literature \cite{Grasso2017}, and it remains difficult to predict where and when any of these will appear during the build process of a given part.
	
	Instead of experimentally tuning the process parameters or part geometry, predictive simulation tries to offer an alternative. The different kinds of modeling approaches for PBFAM can be characterized by the length scales they operate on~\cite{Meier2021, Meier2017}. Mesoscale models are used to analyze the melt pool on length scales from a few powder particles up to one laser scan track \cite{Fuchs2022,Khairallah2014,Korner2011,Markl2015,Meier2021a,Russell2018,Wessels2018,Yan2018}. They can also be used to study 
	the powder recoating process \cite{Herbold2015,Meier2019a,Meier2019, Penny2021}. Microscale models are concerned with the formation of anisotropic metallurgical microstructures during solidification \cite{Gong2015,Lindgren2016,Nitzler2021,Rai2016,Salsi2018,Zhang2013}. In this contribution, we investigate the problem on the macroscale. Since practically relevant geometries are complex, in general, the build process of whole parts needs to be simulated in order to answer questions about the build quality. For this, the term \textit{part-scale} simulation or model is often used in the literature. Virtually all existing part-scale models employ the finite element method (FEM) due to its excellent suitability for thermo(-mechanical) simulations. In this work, we develop an efficient simulation approach for part-scale simulations of the thermal problem. 
	
	The fundamental computational challenge in part-scale simulation lies not so much in the spatial approximation. Although millions of unknowns are necessary to resolve the geometry, state-of-the-art codes and libraries are well-suited to handle this task with mesh adaptivity and parallel processing.   Rather, the challenge lies in the temporal domain. Taking the recent ``AM Bench 2022''~\cite{AMbench2022} build setup as an example, one finds that in order to simulate one of its cantilever specimens with a total scan track of approximately $\num{853} \si{\metre}$ a total of around 44 million time steps (of step size $\num{20} \si{\micro\second}$) are necessary. Put differently, to obtain a solution to this problem within 10 days, one time step may not take longer than $\num{20} \si{\milli\second}$ of wall time. Most classical implementations of FEM models of PBFAM \cite{Hodge2014,Kollmannsberger2018,Riedlbauer2017}, including some of the authors' work \cite{Proell2020,Proell2021}, are suitable for the simulation of a few tracks or layers but do not achieve the level of performance necessary for part-scale simulations. Instead, existing part-scale models use one or more of the following techniques.
	
	A straight-forward approach to part-scale simulations uses a layer-based approach, where whole layers (or parts thereof) are heated at once and the scanning pattern is neglected \cite{Carraturo2020, Dugast2021,Neiva2020,Zhang2018a}. To speed up the simulation further, multiple physical powder layers can be lumped into larger \textit{process layers} \cite{Hodge2016, Zaeh2010}. Typically, these agglomerated models are calibrated with experimental data or resolved single-track or single-layer simulations. Despite the strong simplifications, these models are able to predict, e.g., thermal hot spots or dimensional warping -- but only when calibrated correctly, which can act as a bottleneck or limitation of such approaches.

	In contrast to the literature cited so far, the aim of this contribution is an efficient implementation of PBFAM process simulation with \textit{resolved} scan tracks on hundreds of realistically-sized layers. One prerequisite to enable efficient simulations on that scale is adaptive mesh refinement (AMR). This technique has been employed in various contributions and in different forms \cite{Denlinger2014, ganeriwala2021towards, Li2019, Neiva2019, Olleak2022}. Generally speaking, in AM applications AMR means that the mesh is not static but adapted dynamically over the course of the simulation to be as fine as necessary in the vicinity of the heat source and coarse in regions further away. In addition, the geometry needs to grow to represent the layer deposition in the manufacturing process \cite{Michaleris2014}. Building on top of the \texttt{deal.II} library~\cite{Arndt2022}, its parallel data structures \cite{Bangerth2011, Fehling2022}, and the \texttt{p4est}~\cite{Burstedde2011} library, we develop our own methodology for AMR and growing domains in the targeted PBFAM application.  Our approach is to some extent inspired by a similar strategy, also based on parallel distributed octree meshes, previously presented in \cite{Neiva2019}. While not discussed here, the presented method complements the dual-mortar approach shown in \cite{Proell2021}, which is still relevant for meshing complex geometries.
	
\begin{figure}
		\centering
		\includegraphics[width=.8\linewidth]{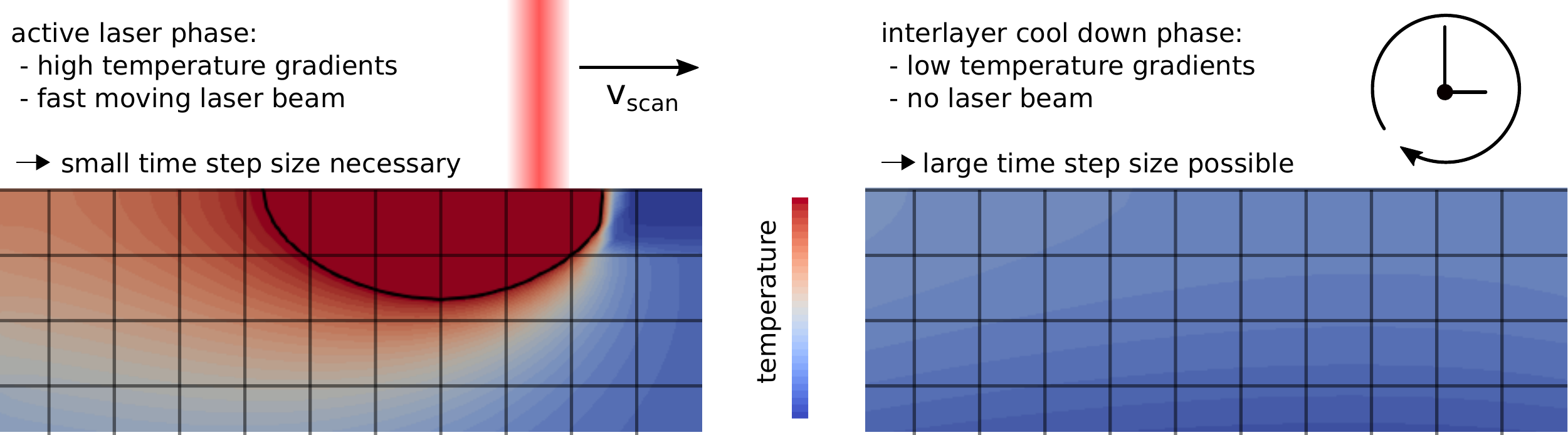}
		\caption{The different phases of the PBFAM build process come with different requirements for the time step size.}
		\label{fig:simulation_phases}
	\end{figure}

	A rarely discussed aspect of efficiency in PBFAM simulations is the choice of time discretization scheme. The PBFAM process can be split into a highly dynamic, active laser phase and a subsequent interlayer cool down phase governed by significantly larger time scales, see Figure~\ref{fig:simulation_phases}. Traditionally, the heat equation is discretized with an unconditionally stable, implicit scheme such as the backward Euler method or generalized trapezoidal scheme. In many applications an implicit scheme seems appropriate as it enables large time steps. This is the case in our application for the cool down phase. Explicit schemes have considerably cheaper evaluation costs per time step and offer better parallel scalability since they can circumvent the assembly of global matrices and the solution of (non)linear systems. However, they are restricted to smaller time steps by a stability limit. It turns out that in the specific scenario of the active laser phase of a scan-resolved PBFAM simulation, the stability limit is not restrictive compared to the time step limitation mandated by the moving heat source. Importantly, the time step limitation due to the moving heat source is required for accuracy and not stability: it holds for explicit and implicit schemes, and thus can be considered an inherent characteristic of the physical problem when modeled in a scan-resolved manner. This consideration has also been stated independently of the authors in the very recent contribution \cite{Essongue2022}. Explicit time stepping has been used for the simulation of PBFAM in \cite{Moran2021}, directed energy deposition in \cite{Mozaffar2019} and wire arc AM in\cite{Huang2020}. In this work, we employ an explicit scheme for the active laser phase and an implicit scheme for the cool down phase. Potential future extensions include a local time stepping scheme or multi-rate time integration \cite{puso2023assessment, Soldner2019} and techniques for temporal decoupling \cite{Hodge2020,Moran2021}.

	For the active laser phase, the evaluation time of FEM integrals becomes the main focus of performance engineering, since a linear system solve can be avoided by using explicit schemes. Some recent contributions in the AM community use GPUs to accelerate the evaluation \cite{Huang2020, Mozaffar2019,Olleak2022}. In \cite{Olleak2022} the authors presented a matrix-free implicit solver for scan-resolved PBFAM simulations.
	
	In this contribution, we will focus on an implementation for CPUs.
	
	Notable other CPU-based implementations of PBFAM models that investigate computational performance make use of AMR and load balancing \cite{Neiva2019}. In addition to AMR, different techniques to scale up the heat source either by elongating it \cite{moreira2022multi}, layer-averaging \cite{zhang20233} or layer-agglomeration \cite{ganeriwala2021towards} are applied in the literature. The present contribution does not use such techniques to stay as close to the physical process as feasible but the methods presented are general enough to incorporate any of these in the future.
	
	To the best of our knowledge, our implementation,  facilitated by the \texttt{deal.II} library and  fast application of FEM operators \cite{Kronbichler2012, Kronbichler2019}, outperforms competing implementations for the thermal PBFAM problem in terms of time to solution. The performance is a result of the  parallel distributed, high-performance implementation of a single time step. The implementation integrates modern hardware features such as vectorized CPU instructions and tries to alleviate  the memory-bound nature (i.e., overall performance is mainly limited by the memory bandwidth rather than the necessary CPU cycles) by the efficient utilization of caches. Another reason for considering CPU-based implementations in this work is that strong scaling is highly relevant for the present application, where CPU-based systems with appropriate tuning often have an edge over GPU systems \cite{Kronbichler2021next}.
	
	The capabilities of the proposed approach are demonstrated on the basis of some challenging examples, the first one being a bridge geometry. Various performance studies show the scalability of the approach on large distributed machines. Finally, and, to the best of the authors' knowledge, for the first time, we present a full scan-resolved simulation of all 312 layers of the NIST AM Bench 2022 cantilever specimen \cite{AMbench2022}. The proposed thermal PBF simulation approach already allows to  assess the influence of scan pattern and part geometry on melt pool shape, overheated zones, zones with residual porosity, which are important indicators for process instabilities. Furthermore, it can serve as a future basis for a thermo-mechanical model to predict thermal distortion and residual stresses or the microstructure in terms of homogenized phase fractions \cite{Nitzler2021} on the scale of real parts.
	
	The remainder of this article is structured as follows: first, we present the mathematical model of the physical process and subsequently derive the spatially and temporally discrete numerical model from it. Next, we discuss aspects of the high-performance implementation with a focus on mesh adaptivity and fast operator evaluation. We present two exemplary numerical simulations of PBFAM on representative geometries and study the performance of the proposed model before we conclude with a short summary and an outlook on future research.

	\section{Mathematical model}
	The present model seeks the solution for the temperature field $T$  in the domain $\Omega$, which is governed by the heat equation:
	\begin{align}
	\label{eq:heat_equation}
	\rho c \pfrac{T}{t} &=  - \nabla \cdot {\vec{q}} + q_\text{vol} ,\quad \vec{q} = -k(T) \nabla T && \text{in } \Omega,
	\end{align}
	with the following parameters: $\rho$ is the density and 
	$c$ is the specific heat capacity of the material. The heat capacity could be used to model the effects of latent heat through an apparent capacity model~\cite{Proell2020}. However, the contribution of latent heat to the overall energy balance is rather small and often neglected in the literature on part-scale AM simulations. For the modeling of a phase-dependent heat conductivity $k$ we briefly summarize the approach from our previous work \cite{Proell2020}. The liquid fraction $g(T)$ is defined as
	\begin{align}
	\label{eq:liquid_fraction}
	g(T) = \begin{cases}
	0, & T < T_s,\\
	\frac{T-T_s}{T_l-T_s}, &T_s \leq T \leq T_l\\
	1, &T > T_l,
	\end{cases}
	\end{align}
	where $T_s$ and $T_l$ are the solidus and liquidus temperature. The time-dependent consolidated fraction
	\begin{align}
	\label{eq:consolidated_fraction}
	r_c(t) = \begin{cases}
	1, & \text{if } r_c(0)=1 \text{ (i.e. initially consolidated)}\\
	\underset{\tilde{t}<t}{\max}\, g(T(\tilde{t})), & \text{if } r_c(0)=0 \text{ (i.e. initially powder)}\\
	\end{cases}.
	\end{align}
    captures the irreversible powder-to-melt phase transition  and allows to set the initial material state.
	From \eqref{eq:liquid_fraction} and \eqref{eq:consolidated_fraction}, the actual fractions of powder ($p$), melt ($m$) and solid ($s$) material are computed as
	\begin{align}
	\label{eq:phase_fractions}
	r_p(r_c) = 1 - r_c,\quad
	r_m(T) = g(T),\quad
	r_s(T,r_c) = r_c -g(T),
	\end{align}
	and finally, the temperature- and history-dependent heat conductivity $k(T,r_c)$ is found:
	\begin{align}
	\label{eq:material_parameter_interp}
	k(T,r_c) = r_p(r_c) k_p + r_m(T) k_m + r_s(T,r_c)  k_s,
	\end{align}
	where $k_p$, $k_s$ and $k_m$ are the single phase parameters. 
		Within each state of the material, all material parameters are fixed, i.e., the single phase problems are linear.
		This choice is made for the sake of simplicity since the focus of this work lies on a HPC implementation of the model rather than on calibration of material data although the implementation presented in this work also supports temperature-dependent parameters.
	Note that the history variable $r_c$ necessitates a proper handling of history data when using mesh adaptivity, e.g., a consistent interpolation of tensor-valued history data \cite{Satheesh2022}.
	
	The volumetric heat source $q_\text{vol}$ models the incident energy from a laser (or electron) beam. In this work, it is given by the following cylindrical model:
	\begin{align}
	\label{eq:heat_soruce_cylindrical}
	q_\text{vol} = \begin{cases} \frac{2 W_\text{eff}}{\pi R^2h_\text{powder}}\exp\left(\frac{-2(\hat{x}^2+\hat{y}^2)}{R^2}\right), & \text{if } 0< \hat{z} < -h_\text{powder} \\
	0, &\text{otherwise}
	\end{cases},
	\end{align}
	which is formulated in a local coordinate system $(\hat{x}, \hat{y}, \hat{z})$ moving along the scan track. The shape in the $xy$-plane is described by a normal distribution with mean $(\hat{x}, \hat{y})=(0,0)$ and standard deviation $\sigma=R/2$. Thus, $R$ can be interpreted as an effective beam radius of the incident energy beam.
	Furthermore, $W_\text{eff}$ is the effective  power, which is reduced compared to the nominal power due to various losses and the material's absorptivity, and $h_\text{powder}$ is the powder layer thickness. The chosen heat source \eqref{eq:heat_soruce_cylindrical} is deliberately kept simple. Other often employed models such as a Gusarov~\cite{Gusarov2009} or Goldak~\cite{Goldak1984} heat source could be easily used instead. In the authors' experience the exact choice does not notably influence the simulation results on the part-scale.
	
	The heat equation \eqref{eq:heat_equation} is completed by the following initial and boundary conditions:
	\begin{align}
		\label{eq:init_condition}
	T &= T_0 && \text{in } \Omega \text{ for } t=0,\\
	\label{eq:bc_dirichlet}
	T &= T_\infty && \text{on } \Gamma_D,\\
	\label{eq:bc_neumann}
	\vec{q} \cdot \vec{n} &= 0 && \text{on } \Gamma_N,\\
	\label{eq:bc_rad_evap}
	\vec{q} \cdot \vec{n} &= q_\text{rad} + q_\text{evap} && \text{on } \Gamma_{RE},\\
	\label{eq:bc_radiation}
& q_\text{rad} =\epsilon \sigma_S(T^4-T_\infty^4), \\
	\label{eq:bc_evaporation}
& q_\text{evap} = \underbrace{0.82 C_P\exp\left[ -C_T \left(\frac{1}{[T]} -\frac{1}{T_v}\right)\right] \sqrt{\frac{C_M}{[T]}}}_{\text{ evaporation mass flux } \dot{m}}\ (h_v +c([T]-T_{h,0})),\ \text{if } [T] > T_v.
	\end{align}
	
	Initially, the whole domain is set to a fixed temperature $T_0$ as stated in \eqref{eq:init_condition}. This is also true for parts of the domain which only become activated at a later stage. The temperature is kept fixed at the ambient temperature $T_\infty$ on the Dirichlet part of the boundary $\Gamma_D$ at the bottom of the baseplate \eqref{eq:bc_dirichlet}. Both a radiation and evaporation condition \eqref{eq:bc_rad_evap}  are applied on the free surface $\Gamma_{RE}$ at the top of the built part. Inclusion of a convection boundary condition would be straight-forward but not done in this work as the influence is considered small as compared to radiation and evaporation. The remaining part of the boundary $\Gamma_N$ is modeled as thermally insulating \eqref{eq:bc_neumann}. These conditions include the following constants and parameters:
	For the radiation term \eqref{eq:bc_radiation}, $\epsilon$ is the emissivity and $\sigma_S$ the Stefan-Boltzmann constant.
	For the evaporation condition based on \cite{anisimov1995instabilities,Meier2021a}, $C_P = 0.54p_a$ is a factor with the dimension of pressure computed from the atmospheric pressure $p_a$ and $C_T \approx \bar{h}_v/R$ a factor with the dimension of temperature computed from the molar latent heat of evaporation $\bar{h}_v$ and the molar gas constant $R$. Moreover,  $T_v$ is the boiling temperature, $h_v$ the specific latent heat of evaporation and $T_{h,0}$ is a reference temperature for the enthalpy calculation. The constant $C_M = M/(2\pi R)$ is computed from the molar mass $M$ and the molar gas constant $R$. Overall this leads to an expression for the heat flux $q_\text{evap}$ from evaporation that consists of an evaporative mass flux $\dot{m}$ multiplied by a specific enthalpy. To avoid numerical issues with the strong nonlinearity in the evaporation term \eqref{eq:bc_evaporation}, the temperature $[T]$ used for its evaluation is limited to a maximum value $T_\text{max} > T_v$ by setting $[T] = \min(T, T_\text{max})$. In this work, we choose $T_\text{max} = T_v + 1000 \si{\kelvin}$. This choice does not influence the overall results and leads to a robust numerical scheme.

	\section{Numerical discretization and solution schemes}
	
	\subsection{Weak form and spatial discretization}
	
	In order to solve the heat equation \eqref{eq:heat_equation} numerically we employ a finite element (FE) discretization for the spatial dimension. First, the heat equation is multiplied with a test function $v$ and the diffusive term is integrated by parts, yielding
	\begin{align}
	\label{eq:weak_form}
	\left(v, \rho c \pfrac{T}{t}\right)_\Omega =  \left(\nabla v, \vec{q}\right)_\Omega - \left(v, \vec{q}\cdot\vec{n}\right)_{\Gamma_{RE}} + \left(v, q_\text{vol}\right)_\Omega,
	\end{align}
	where $(a,b)_\square := \int_\square ab$. The weak form \eqref{eq:weak_form} is
	equivalent to the strong form \eqref{eq:heat_equation} if the test
	function is chosen from the weighting space $\mathcal{W} = \lbrace v \in H_1 (\Omega) : v = 0 \text{ on } \Gamma_D \rbrace$ and the solution function is chosen from the trial
	space $\mathcal{V} = \lbrace T \in H_1 (\Omega) : T = T_\infty \text{ on } \Gamma_D\rbrace$ , where $H_1 (\Omega)$ is the Sobolev space containing functions
	with square-integrable first derivatives. The solution and test functions are discretized in space
	based on a (continuous) Bubnov-Galerkin ansatz:
	\begin{align}
	\label{eq:bubnov_galerkin}
	T_h(\vec{x},t) = \sum \varphi_j(\vec{x})T_j(t), \quad v_h(\vec{x},t) = \sum \varphi_j(\vec{x})v_j(t),
	\end{align}
	where $\vec{x}$ is the spatial coordinate, $\varphi_j(\vec{x})$ are the space-dependent shape functions used for solution and test functions. The discrete degrees of freedom (DoFs) $T_j(t)$ and $v_j(t)$ only depend on time. In this work, we exclusively use first-order Lagrange polynomials but the implementation supports higher order functions as well. After inserting the spatial discretization \eqref{eq:bubnov_galerkin} into the weak form \eqref{eq:weak_form} we obtain the following semi-discrete problem:
	\begin{align}
	\label{eq:semi-discrete}
	\vec{C}\dot{\vec{T}} = \vec{f}(\vec{T}) = \vec{f}_\text{diff}(\vec{T}) + \vec{f}_\text{RE}(\vec{T}) + \vec{f}_\text{vol},
	\end{align}
	where $\vec{C}$ is a capacity matrix, $\vec{T}$ and $\dot{\vec{T}}$ are the global vectors of nodal temperatures and their time derivatives and $\vec{f}(\vec{T})$ is composed of the nonlinear diffusive term $\vec{f}_\text{diff}(\vec{T})$ as well as the boundary term $ \vec{f}_\text{RE}(\vec{T})$ and source term $\vec{f}_\text{vol}$. These terms are given in the same order as their equivalent weak form contributions in \eqref{eq:weak_form}.
	
	\subsection{Time integration and solution procedure}
	
	In this contribution, an implicit and an explicit time integration scheme are combined. When a layer is scanned, the explicit time integration scheme is used, while the interlayer cool down phase is simulated with the implicit time integration scheme. Since the scanning phase requires most of the computational time, we mainly tune the performance of the explicit scheme, as discussed in the next section.
	
	\paragraph{Explicit scheme} For the active laser phase we apply the forward Euler scheme to \eqref{eq:semi-discrete}:
	\begin{align}
	\label{eq:forward_euler}
	\vec{T}_{n+1} = \vec{T}_n +\Delta t \tilde{\vec{C}}^{-1}\vec{f}(\vec{T}_{n}),
	\end{align}
	where the consistent capacity matrix $\vec{C}$ is replaced with a lumped, diagonal variant~\cite{Kormann2016},
	\begin{align}
	\tilde{C}_{ii} = \sum_j C_{ij},
	\end{align}
	which is trivially invertible. In the implementation, the diagonal matrix can be precomputed and stored as a vector such that its application becomes a simple scaling operation. The computationally most challenging task in \eqref{eq:forward_euler} is the efficient evaluation of $\vec{f}(\vec{T}_{n})$. Note that any other explicit time stepping scheme could be used as well. We found the explicit Euler scheme to provide sufficient accuracy for the small time steps sizes (demanded by the moving heat source).
	
	Explicit time integration schemes are not unconditionally stable. In order to find the largest stable time step for the explicit Euler scheme, we replace the nonlinear function in \eqref{eq:forward_euler} with a linearized version which only considers the critical diffusive term $\vec{f}_\text{diff} \approx \vec{K}_\text{diff}\vec{T}$:
	\begin{align}
	\label{eq:forward_euler_taylor}
	\vec{T}_{n+1} \approx \vec{T}_n +\Delta t \tilde{\vec{C}}^{-1}  \vec{K}_\text{diff}\vec{T}_n
	= \underbrace{\left(\vec{I} + \Delta t \tilde{\vec{C}}^{-1} \vec{K}_\text{diff} \right)}_{=: \vec{A}} \vec{T}_n = \vec{A}^n\vec{T}_0,
	\end{align}
	where $\vec{I}$ is the identity matrix. The approximation performed in \eqref{eq:forward_euler_taylor} only neglects the nonlinearity in the heat conductivity which is limited to the small phase change interval $\lbrack T_s, T_l\rbrack$. Since the matrix $\vec{A}$ is repeatedly applied to the temperature vector, its spectral radius must be $\rho(\vec{A}) \leq 1$, i.e., its largest absolute eigenvalue needs to be smaller than 1. After some rearrangement one finds for the critical time step:
	\begin{align}
	\label{eq:explicit_time_step}
	\Delta t \leq \frac{2}{\rho
		\left(\tilde{\vec{C}}^{-1} \vec{K}_\text{diff}\right)}.
	\end{align}
	A more detailed discussion of stability limits in the context of explicit time integration for PBFAM problems can be found in \cite{Essongue2022}.
	In addition to \eqref{eq:explicit_time_step}, the admissible time step size is also limited by the velocity $v_\text{scan}$ of the moving heat source: we do not want the heat source to travel further than the radius $R$ of the laser beam in one time step and therefore require:
	\begin{align}
	\label{eq:moving_heat_source_condition}
	\Delta t \leq  \frac{R}{v_\text{scan}}.
	\end{align}
	It is crucial to realize that \eqref{eq:moving_heat_source_condition} is required to achieve a continuous melt track in a scan-resolved simulation. Thus, this restriction also holds for an implicit scheme which might allow much larger time steps from a pure stability perspective. With our choice of heat source model \eqref{eq:heat_soruce_cylindrical}, if one were to use a larger time step than \eqref{eq:moving_heat_source_condition} allows, the melt track would break up into disjoint segments. Note that this restriction could be weakened by the use of elongated line heat sources of equivalent energy \cite{chiumenti2017numerical, irwin2016line} but since such an approach requires calibration, we do not follow it here. Together \eqref{eq:explicit_time_step} and \eqref{eq:moving_heat_source_condition} form a combined criterion for the maximum time step size:
	\begin{align}
	\label{eq:time_step_combined}
	\Delta t \leq \min \left\lbrace  \frac{R}{v_\text{scan}},\ \frac{2}{\rho
		\left(\tilde{\vec{C}}^{-1} \vec{K}_\text{diff}\right)}\right\rbrace.
	\end{align}
	Estimation of the spectral radius is rather expensive; consequently,  the stability criterion should only be evaluated in the setup phase of a simulation. In the numerical examples we found it sufficient to only evaluate \eqref{eq:time_step_combined} once for the critical values in a given set of parameters since the characteristics involved in the criterion do not change over layers. For the parameters used in the numerical examples, we found the accuracy criterion \eqref{eq:moving_heat_source_condition} (which is independent of the time integration scheme) to be around 5-10 times smaller than the stability criterion for the explicit scheme \eqref{eq:explicit_time_step}. Evaluation of an explicit time step will be faster than the (iterative) solution of a linear system arising from an implicit scheme. Therefore, we conclude that an explicit scheme is superior for the simulation of the active laser phase.

	\paragraph{Implicit scheme}
	 The following implicit scheme can be applied to \eqref{eq:semi-discrete}:
	\begin{align}
	\label{eq:backward_euler_nonlinear}
	\vec{r} := \frac{1}{\Delta t} \vec{C}(\vec{T}_{n+1}-\vec{T}_n)-\vec{f}_\text{diff}(\vec{T}_{n+1}) -\vec{f}_\text{RE}(\vec{T}_n) -\vec{f}_\text{vol} = \vec{0}.
	\end{align}
	Note that the radiation and evaporation boundary terms are evaluated for the previous time step. Since the implicit scheme is only used for the interlayer cool down phase (which does not exhibit high temperature gradients and rates compared to the active laser phase), this choice has no influence on the robustness and accuracy of the solution as verified by our investigations. To solve the nonlinear system of equations in residual form  \eqref{eq:backward_euler_nonlinear} for the unknown temperatures $\vec{T}_{n+1}$ a Newton-Raphson scheme is used:
	\begin{align}
	\label{eq:Newton_1}
	\underbrace{\left(\frac{1}{\Delta t} \vec{C} - \pfrac{\vec{f}_\text{diff}}{\vec{T}_{n+1}}({\vec{T}_{n+1}^i})\right)}_{\vec{J}_{\vec{r},n}^i}\Delta \vec{T}_{n+1}^{i+1}
	&= -\underbrace{\left(\frac{1}{\Delta t} \vec{C}(\vec{T}_{n+1}-\vec{T}_n)-\vec{f}_\text{diff}(\vec{T}_{n+1}) -\vec{f}_\text{RE}(\vec{T}_n) -\vec{f}_\text{vol}\right)}_{\vec{r}_n^i},\\
	\label{eq:Newton_2}
	\vec{T}_{n+1}^{i+1} &= \vec{T}_{n+1}^{i}+\Delta \vec{T}_{n+1}^{i+1},
	\end{align}
	where $\vec{J}_{\vec{r},n}^i = \pfrac{\vec{r}}{\vec{T}_{n+1}}({\vec{T}_{n+1}^i})$ is the Jacobian of the residual evaluated at the current temperature iterate $\vec{T}_{n+1}^i$. This iterative scheme \eqref{eq:Newton_1}--\eqref{eq:Newton_2} is applied until convergence of the residual \eqref{eq:backward_euler_nonlinear} is achieved up to a prescribed tolerance. In this contribution, we use the implicit scheme with matrix-free evaluation in combination with an infrequently updated preconditioner (incomplete LU-factorization) for the linear solver only to solve the large time steps in the interlayer cool down phase. While technically, this scheme could be used to obtain a stable result for large time step sizes during the active laser phase, this is not done in this work for the accuracy reasons stated in \eqref{eq:moving_heat_source_condition}.  In Appendix~\ref{sec:temporal_convergence}, we demonstrate temporal convergence and robustness of the combined time integration scheme.
	
	\section{High-performance implementation}
	
	The numerical model summarized in the previous section is implemented in an in-house research code based on the \texttt{deal.II} finite element library \cite{Arndt2022}. General-purpose functionality developed in the context of this work has been contributed to the main \texttt{deal.II} repository. 
	In this section, we discuss implementation details. Note, that in this section the term \textit{process} refers to either an \textit{evaluation process} (the act of computing a result, an algorithm) or a \textit{computer process} (a program instance executed by the CPU). It does not refer to the simulated AM process.
	
	\subsection{Mesh adaptivity and layer activation}
	\label{sec:adaptivity_layers}
	
\begin{figure}
		\centering
		\includegraphics[width=\linewidth]{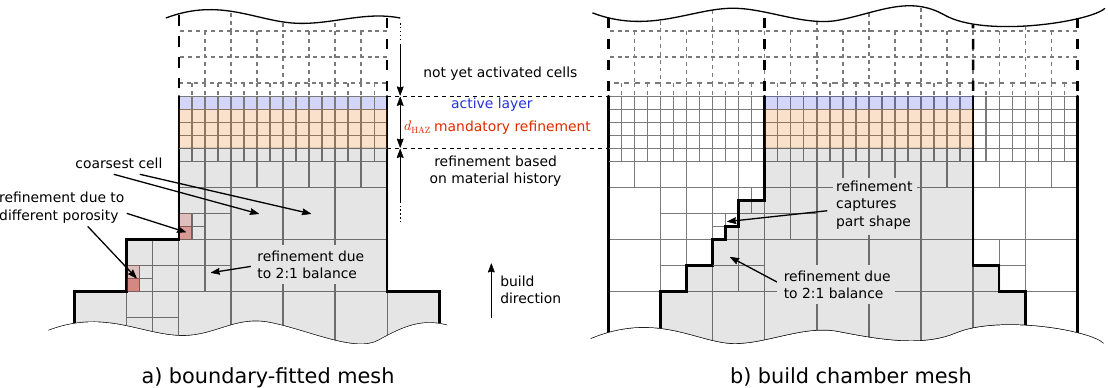}
		\caption{Adaptive mesh refinement concept applied in AM context. The proposed framework can work with a) boundary-fitted meshes and b) build chamber meshes.}
		\label{fig:amr_mesh_concept}
	\end{figure}

	Adaptive meshes can enable large savings in CPU time and memory usage and ultimately speed up the solution time considerably. For the present AM application, we can predict \textit{a priori} when and where a fine mesh is needed without the need for an \textit{a posteriori} error estimator. Therefore, we suggest the following procedure for mesh adaptivity and activation of new layers as illustrated in Fig. \ref{fig:amr_mesh_concept}.
	
	The complete part geometry is created in the beginning and meshed with a coarse mesh consisting of hexahedral cells of uniform edge length. In the current implementation, the edge length $h_\text{coarse}$ of the coarse mesh should be related to the desired powder layer height $h_\text{powder}$ via
	\begin{align}
	\label{eq:coarse_fine_relation}
	h_\text{coarse} = 2^{n_\text{refine}} \cdot h_\text{powder},
	\end{align}
	such that $n_\text{refine}$ is the number of necessary isotropic refinements (by subdivision) of a coarse cell to obtain cells of the same height as the powder layer $h_\text{powder}$. This procedure allows for a straight-forward application of new powder layers as the boundaries of the powder layer always coincide with cell boundaries. The coarse mesh size $h_\text{coarse}$ (or, equivalently, the number of refinements $n_\text{refine}$) should be chosen as large as possible to realize the largest computational savings. This approach works well for the geometries investigated so far and allows for a simple transfer of data across matching mesh hierarchies. Should more complex meshes be necessary, one could relax the constraint \eqref{eq:coarse_fine_relation} and use more general transfer operations (e.g., based on mortar meshtying schemes) between potentially non-matching meshes as demonstrated in our previous work~\cite{Proell2021}.
	
	For mesh generation, one option is a coarse mesh which directly represents the final part geometry as a boundary-fitted voxel mesh. In this case, no surrounding powder is modeled, i.e., the boundaries of the coarse mesh are the boundaries of the part, see Fig.~\ref{fig:amr_mesh_concept}a). This approach is justified by the very low thermal conductivity of the powder, which  can be approximated by a thermally isolating boundary condition. Alternatively, the coarse mesh can represent a powder-filled build chamber and the boundaries of the coarse mesh can be interpreted as the boundaries of the build chamber. In this case, the final part geometry is defined implicitly from the consolidation status of every material point, see Fig.~\ref{fig:amr_mesh_concept}b). Both geometry descriptions are possible within our framework and they both have specific advantages: the boundary-fitted coarse mesh allows to coarsen most cells that are far away from the currently scanned layer but generation of such a coarse mesh that still represents the part shape accurately can be cumbersome. On the other hand, the non-fitted build chamber mesh is trivial to generate as the domain will be a cuboid. However, there is a certain overhead in areas that are meshed, although they are not necessary for the representation of the final part shape and, in addition, cells close to the implicit part boundary need to stay refined throughout the simulation to capture the final part shape. We present examples utilizing both meshing approaches. Note that the approximation of the part geometry can also be realized via other methods, e.g., the finite cell method \cite{Carraturo2020} or CutFEM \cite{burman2015cutfem,Schott2019}. 
	
	Whenever a new powder layer is added, refinement, coarsening and activation of cells takes place according to the following rules: cells are refined such that the currently scanned, top-most layer is represented with a desired number of cells over the layer height. Note that all cells in the current layer are refined upon activation, regardless of when or if the laser reaches them. This avoids the computational effort for frequent remeshing within a layer at the cost of slightly more DoFs.
	Cells in the heat affected zone (HAZ) -- a few layers below the current layer -- also stay refined. Cells which have a distance greater than $d_\text{HAZ}$ (which we choose as $d_\text{HAZ}=4h_\text{powder}$) from the current layer \textit{may} be coarsened with the following restriction: for any set of eight cells, which are octants of a previously subdivided parent cell, coarsening only takes place if the material history state -- the consolidation state $r_c$ defined in \eqref{eq:consolidated_fraction} -- across this set lies above a threshold of $r_\text{coarsen} = 0.9$. This restriction ensures that potential porosity defects are not smoothed out over neighboring cells of full density and that the part boundaries stay refined up to the necessary level when using a build chamber mesh. It should be mentioned that the employed \texttt{p4est} library enforces a 2:1 balance between refinement levels of neighboring cells, i.e. for any two neighboring cells the refinement level may differ by at most one.
	
	To save computational resources, all cells that lie above the currently active layer are inactive and coarsened as much as possible. \textit{No} DoFs are assigned to them and they need \textit{not} be evaluated; thus, they implicitly represent void.  When discussing the parallel distribution of the cells, two aspects should be distinguished: first, how many processes should be used in total, and second, how to distribute the cells among the processes.
	In this work, the maximum number of unknowns determines the allocated number of processes.  No dynamic resource allocation takes place in the current implementation and the same number of processes is used throughout the entire simulation. Due to the growing geometry, it seems resaonable to allocate more and more processes as the simulation progresses. However, such a  dynamic resource allocation  scheme is expected to save resources but not necessarily speed up the overall simulation, since we are not typically limited by the spatially distributed scale of the problem but rather by the temporal scale. In practical AM simulation setups, the number of CPUs will be increased until the scaling limit is reached, i.e., until no further speedup can be achieved due to an increasing communication overhead and the parallel efficiency drops below an acceptable threshold. Next, regarding the distribution of cells among all processes, the inactive cells are not weighted differently compared to active cells. Instead, all parallel processes receive roughly the same number of cells regardless of the computational effort within the cell. This implies that for large process counts some processes will not have any work in the initial layers. A first attempt at a weighted redistribution of the active cells, that tries to utilize more processes for actual work, did not result in a noticeable speedup, possibly due to non-negligible communication latency
	for such configurations  introduced by the specific Z-curve ordering of cells \cite{Burstedde2011}.
	 Thus, detailed investigations of these aspects are left for future work.  The interested reader is referred to \cite{Neiva2019}, where the performance of a similar AMR strategy was also rather insensitive to a weighted partitioning.

	In the context of AM problems, the new \texttt{deal.II} class \texttt{FieldTransfer} was developed which allows to transfer global state vectors between meshes after coarsening and refinement in the presence of inactive cells (that do not have any DoFs). This class was initially developed by the authors of \cite{Kim2022} and improved by us; it is thus applicable to various types of growing domain problems encountered when modeling different AM processes.
	
	\subsection{Fast operator evaluation}
	
	In the discrete problem statement in \eqref{eq:forward_euler} and \eqref{eq:Newton_1} we need to evaluate volume and boundary face integrals. Their efficient implementation uses the same techniques and, for the sake of brevity, we demonstrate the fast operator evaluation on the diffusive term in \eqref{eq:weak_form}, which is a crucial term both in the explicit and implicit formulation of our problem. This global integral over the heat flux $\vec{q}$ can be transformed into a sum of element-level integrals, which are assembled into a global vector $\vec{f}_\text{diff}$:
	\begin{align}
	(\nabla_{\vec{x}} v, \vec{q})_{\Omega} = \sum_e (\nabla_{\vec{x}} v_e, -k(T_e) \nabla_{\vec{x}} T_e)_{\Omega_e}
	= \sum_e v_{e,i}\underbrace{(\nabla_{\vec{x}} N_i, -k(T_e) \nabla_{\vec{x}} T_e)_{\Omega_e}}_{f_{e,i,\text{diff}}} = \sum_e \vec{v}_e^T\vec{f}_{e,\text{diff}} =:  \vec{v}^T\vec{f}_{\text{diff}},
	\end{align}
	where the index $e$ indicates a quantity restricted to a single element and the operator $\nabla_{\vec{x}}$ is used to represent the gradient in physical space. Following \cite{Kronbichler2012}, we now break down the computation further:
	\begin{align}
	\label{eq:diffusive_term_evaluation}
	f_{e,i, \text{diff}}
	&= \int_{\Omega_e}(\nabla_{\vec{x}} N_i)^{T} (-k(T_e) \nabla_{\vec{x}} T_e)\, \dd \vec{x} \nonumber\\ 
	&= \int_{\Omega_e}(\vec{J}_e^{-T}\nabla_{\vec{\xi}}N_i)^T (-k(N_aT_{e,a}) (\vec{J}_e^{-T}\nabla_{\vec{\xi}} N_b)T_{e,b}) \det \vec{J}_e\, \dd \vec{\xi} \nonumber \\
	&\approx \sum_q \underbrace{(\nabla_{\vec{\xi}}N_{i})^T}_{(\vec{S}_\text{grad}^T)_{ic}}\underbrace{\vec{J}_{e}^{-1}w (\det \vec{J}_{e}) \vec{J}_{e}^{-T} (-k\underbrace{(N_{a}T_{e,a})}_{\vec{S}_\text{val}\vec{T}_e})}_{:= (\vec{D}_e)_{cd}}\underbrace{(\nabla_{\vec{\xi}} N_{b})T_{e,b}}_{(\vec{S}_\text{grad}\vec{T}_e)_{d}}.
	\end{align}
	In the second line, the element integral is transformed from the physical space (parametrized by coordinates $\vec{x}$) into reference space (parametrized by coordinate $\vec{\xi}$), thereby introducing the Jacobian mapping $\vec{J}_e$ between these spaces. Also, the temperature $T_e$ is discretized in space with the element-wise shape functions $N_a$ which correspond to the global spatial discretization introduced in \eqref{eq:bubnov_galerkin}. To shorten the notation, we employ the Einstein summation convention. In the third line, the integration is replaced by a weighted sum according to a numerical quadrature rule. Here, the shape functions $N_i$, their derivatives $\nabla_{\vec{\xi}}N_i$ and the Jacobian mapping $\vec{J}_e$ are all evaluated at those quadrature points and $w$ is the quadrature weight. Although the final equation \eqref{eq:diffusive_term_evaluation} is lengthy, it illustrates the sequential nature of the element-level evaluation: first, we obtain the values and gradients of the temperature at the quadrature points through the interpolation $\vec{S}_\text{val}$ and $\vec{S}_\text{grad}$. In the implementation, we use a technique known as sum-factorization, which has been established in the spectral element community \cite{Deville2003,Melenk2001,Orszag1980} and is available via \texttt{deal.II} \cite{Kronbichler2012,Kronbichler2019}. Sum-factorization is especially beneficial for tensor-product shape functions of polynomial degree two and higher. However, in the present case of linear shape functions it provides more opportunities for the compiler to optimize code, e.g., through register blocking. Given the values and gradients of $T$, all physics-related operations happen on quadrature point level inside $\vec{D}_e$. In this example, we compute the heat flux from the nonlinear conductivity and thus need to evaluate the value of the temperature via $\vec{S}_\text{val}$. Finally, these quadrature point contributions are multiplied with  $\vec{S}_\text{grad}^T$ (the shape gradients resulting from the test function) and summed up.
	
	There is one missing link for the complete picture, namely the relation between element-level quantities and global quantities. For this purpose, we introduce a gather operation $\vec{G}_e$ which extracts local DoFs from a global vector via
	\begin{align}
	\vec{T}_e = \vec{G}_e\vec{T}.
	\end{align}
	The transpose of the gather operations $\vec{G}_e^T$ scatters an element contribution back into a global vector such that we can write for the whole evaluation process:
	\begin{align}
	\label{eq:vector_evaluation}
	\vec{f}_\text{diff} = \sum_e \vec{G}_e^T\vec{S}_\text{grad}^T\vec{D}_e\vec{S}_\text{grad}\vec{G}_e\vec{T}.
	\end{align}
	Note that internally $\vec{D}_e$ also requires the values of the temperature at quadrature points (computed via $\vec{S}_\text{val}\vec{G}_e\vec{T}$), since the conductivity $k$ depends on it, see \eqref{eq:diffusive_term_evaluation}. Equation \eqref{eq:vector_evaluation} demonstrates how to assemble a global vector from cell-wise contributions.
	Interestingly, an equivalent strategy can be applied to compute a matrix-vector product without assembling the matrix first. As an example, we look at the matrix-vector product on the left-hand side in \eqref{eq:Newton_1}. Following the same steps as before and taking into account the definition of $\vec{C}$ and $\vec{f}_\text{diff}$ according to \eqref{eq:weak_form} and \eqref{eq:semi-discrete}, we arrive at the following evaluation process:
	\begin{align}
	\label{eq:matrix_free_eval}
	\vec{J}_{\vec{r},n}^i \Delta \vec{T}_{n+1}^{i+1} &= \sum_e \vec{G}_e^T\vec{S}_\text{val}^T\left[\frac{\rho c}{\Delta t} \vec{I}\right]  \vec{S}_\text{val}\vec{G}_e\Delta\vec{T}_{n+1}^{i+1} - \sum_e \vec{G}_e^T\vec{S}_\text{grad}^T\left[\vec{D}_e \vec{S}_\text{grad} + \vec{L}_e \vec{S}_\text{val}\right] \vec{G}_e\Delta\vec{T}_{n+1}^{i+1},\\
	&\text{where } (\vec{L}_e)_c = \vec{J}_{e}^{-1}w \det \vec{J}_{e} \vec{J}_{e}^{-T} \left[-\pfrac{k}{T}\underbrace{(N_{a}T_{e,a})}_{\vec{S}_\text{val}\vec{T}_e} \underbrace{\nabla_{\vec{\xi}} N_{b}T_{e,b}}_{(\vec{S}_\text{grad}\vec{T}_e)_{c}} \right]
	\end{align}
	Equation\eqref{eq:matrix_free_eval} illustrates another feature of the evaluation process: in contrast to classical FEM implementations no global matrix is assembled. Instead, we directly compute the effect of the Jacobian $\vec{J}_{\vec{r},n}^i$ on the vector $\Delta \vec{T}_{n+1}^{i+1}$ and obtain the result as another global vector. Thus, the global Jacobian needed in the implicit time integration scheme \eqref{eq:Newton_1} does not need to be assembled explicitly. Due to the aforementioned property the whole evaluation process is often termed \textit{matrix-free} evaluation. Essentially, this process moves the bottleneck from the memory transfer of matrix entries (matrix-based implementation) to a more intense calculation (repeated calculation of \eqref{eq:matrix_free_eval} for every matrix-vector product). This trade-off is often preferable in modern hardware and especially pronounced for higher-order shape functions \cite{Kronbichler2018}.
	Similar to \cite{Schoeder2018} we use the presented algorithm for explicit time integration in \eqref{eq:vector_evaluation} (which is inherently matrix-free) due to its mature and optimized implementation in \texttt{deal.II} \cite{Kronbichler2012,Kronbichler2019}.
	
	The process illustrated so far is very general and does not require any assumptions on the integrated term. In fact, the diffusive term includes nonlinear and history-dependent behavior which can be evaluated at each quadrature point. Furthermore, the procedure transfers seamlessly to (boundary) faces and integrals over them. The different steps of the evaluation still allow for a multitude of optimizations that can be chosen under specific circumstances: such optimization strategies and their trade-offs are discussed in detail in \cite{Kronbichler2019}. Some algorithmic aspects 
	of special relevance to this article are presented in the following subsections.

	\begin{remark}
		The various operations that make up the whole evaluation process described in this section have been written in a matrix-vector product notation to better illustrate the process. This has only been done for readability, as they are not necessarily implemented with matrix-vector products as this does not lead to the most efficient implementation as discussed in \cite{Kronbichler2012,Kronbichler2019}.
	\end{remark}
	
	\begin{remark}
		Note that we can still assemble a global matrix via the matrix-free evaluation in \eqref{eq:matrix_free_eval} (e.g. to construct a reusable matrix-based preconditioner) by multiplying with each unit vector and assembling the result vectors as the columns of a global matrix.
	\end{remark}
	\subsubsection{Vectorization over cell batches}
	
	\begin{figure}
		\centering
		\includegraphics[width=.8\linewidth]{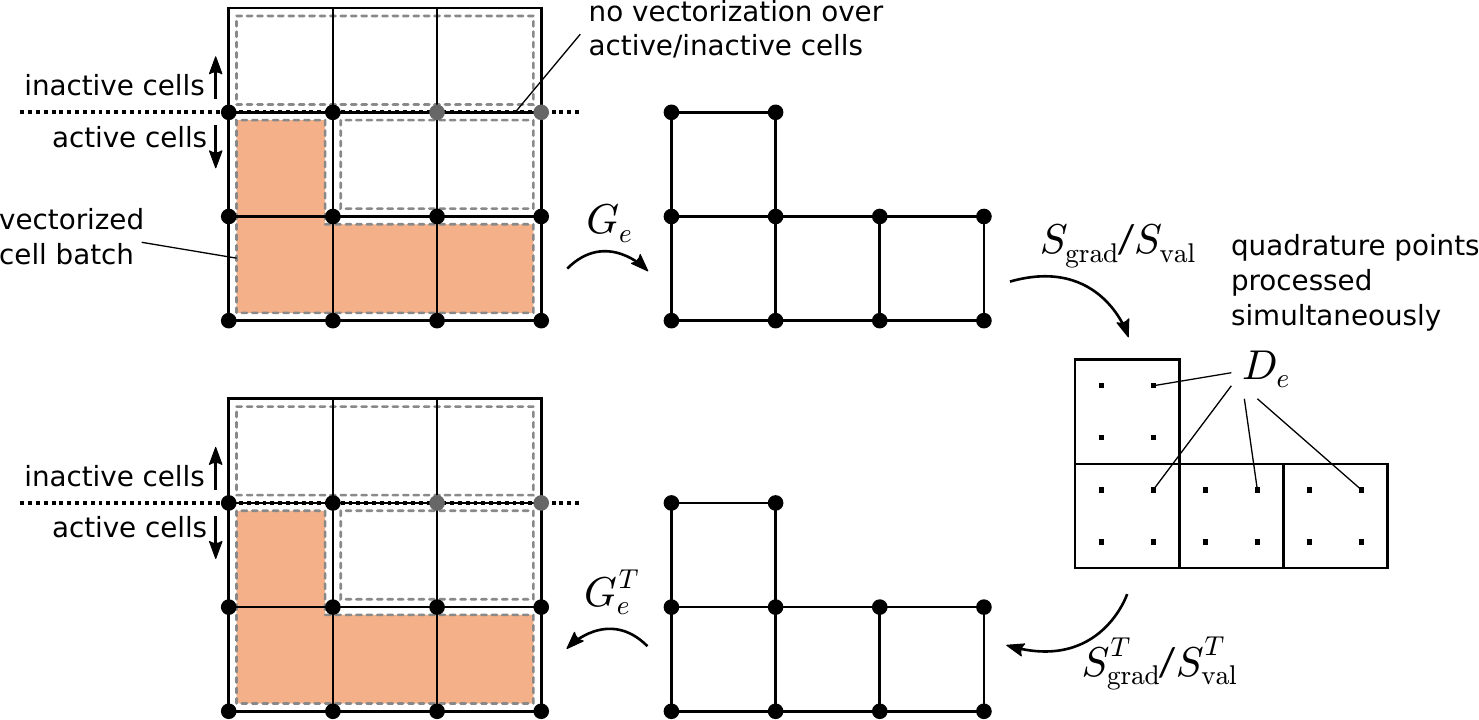}
		\caption{Illustration of fast operator evaluation on a vectorized cell batch. The operation $\vec{D}_e$ is simultaneously performed on the same relative quadrature points in a vectorized cell batch with vectorized CPU instructions.}
		\label{fig:vectorization}
	\end{figure}
	
	Modern CPUs come with single-instruction-multiple-data (SIMD) capabilities for basic arithmetic operations and loading and storing of memory. For instance, the AVX2 instruction set architecture allows to perform a single operation on 4 different values of type \texttt{double} at the same time. The more recent AVX512 instruction set even allows for 8 values of type \texttt{double}. Although optimizing compilers will try to identify loops that benefit from a vectorization of instructions on their own, the effectiveness of compiler optimizations depends on the specific implementation of the FEM model.  
	If one were to rely only on the compiler for auto-vectorization and if the quadrature point operation is simple enough, the compiler might vectorize the operations done at a single quadrature point or reorder operations to process operations at different quadrature points of the same cell simultaneously.
	However, data dependencies and non-unit-stride access often prevent this automatic approach from utilizing SIMD effectively, and much better performance is possible by processing the operations of different cells in each vector lane, as demonstrated in \cite{Kronbichler2019}. This outer-loop vectorization ensures that all SIMD lanes can be filled with useful work, avoiding remainder loops or mask operations that a compiler would generate for auto-vectorization.
	In this contribution, intrinsics-based explicit vectorization is facilitated by abstractions of the \texttt{deal.II} library, without leaking the details of the loop for vectorization to the user code.

	Figure \ref{fig:vectorization} illustrates how the operator evaluation described in the last section works on a batch of cells simultaneously. The number of cells in a batch is determined by the hardware which supports a number of lanes $n_\text{lanes}$. In the illustration the vectorization width is 4. 
	In our PBFAM application, vectorization is performed separately over the activated cells in/below the current layer and the inactive void cells above (where we do not evaluate the weak form but perform post-processing operations for visualization). The vectorization concept also extends to the evaluation of integrals on (boundary) faces of cells. 
	
	For the full performance of the vectorized instructions, the nonlinear history behavior of the material is fully vectorized. For linear shape functions, the quadrature point operation constitutes more than half of the arithmetic work. The original formulation \cite{Proell2021} contains branching conditions (e.g. in the liquid fraction evaluation \eqref{eq:liquid_fraction}) which is often implemented with \texttt{if}-statements in unvectorized codes. These statements can be reformulated via masking operations. For a value $a$ let us denote its vectorized version as $\check{a}$ and access to entry $i$ by $\check{a}[i]$.  For instance, we can define various masks that filter for temperatures in a given interval:
	
	\begin{align}
	\label{eq:mask_example_1}
	\check{M}_{T<T_s} &= \check{\text{mask}}_<(\check{T}, \check{T}_s, \check{1}, \check{0}), \\
	\label{eq:mask_example_2}
	\check{M}_{T_s<T<T_l} &= \check{\text{mask}}_>(\check{T}, \check{T}_s, \check{1}, \check{0}) \cdot \check{\text{mask}}_<(\check{T}, \check{T}_l, \check{1}, \check{0}), \\
	\label{eq:mask_example_3}
	\check{M}_{T>T_l} &= \check{\text{mask}}_>(\check{T}, \check{T}_l, \check{1}, \check{0}), \\
	&\text{where}\ \check{\text{mask}}_{\square}(\check{a},\check{b},\check{t},\check{f})[i] = \begin{cases}
	\check{t}[i],\quad \text{if}\ \check{a}[i]\ \square\ \check{b}[i], \\
	\label{eq:mask_function}
	\check{f}[i],\quad \text{otherwise}
	\end{cases} \text{ for } 0 \leq i < n_\text{lanes}.
	\end{align}
	Each of these masks $\check{M}_C$ contains a one in every lane where the condition $C$ in the subscript is true, and a zero otherwise. 
	To get a filtering effect, the masks can be multiplied with any quantity that should only be considered when the condition is true. The C++ language supports operator overloading such that the vectorized result of the $\check{\text{mask}}$-function and the final masks $\check{M}_C$ support the usual arithmetic operations. Note that the $\check{\text{mask}}$-function is implemented with intrinsic SIMD-calls specific to a given architecture and \eqref{eq:mask_function} only documents its behavior. In our framework, the complete material behavior and weak form are consistently implemented on vectorized data types. For instance, a vectorized version of the liquid fraction \eqref{eq:liquid_fraction} can be written with the help of the masks \eqref{eq:mask_example_1}--\eqref{eq:mask_example_3} as
	\begin{align}
	\check{g}(\check{T}) = \check{M}_{T<T_s} \cdot \check{0} + 	\check{M}_{T_s<T<T_l} \cdot \frac{\check{T}-\check{T_s}}{\check{T}_l -\check{T}_s} + 	\check{M}_{T>T_l} \cdot \check{1}.
	\end{align}
	This illustrates that we can compute the liquid fraction for $n_\text{lanes}$ quadrature points simultaneously, while the computational cost is comparable to a single quadrature point evaluation in the unvectorized implementation. The liquid fraction $\check{g}$ is used to compute the consolidated fraction history variable $\check{r}_c$ according to \eqref{eq:consolidated_fraction} which needs to be stored for every quadrature point and is transferred to a new mesh upon coarsening.

	\subsubsection{Further aspects}
	So far, the discussion of performance has been about the cell-local operation of a single operator evaluation only. However, we would like to note that we do not evaluate the terms of the weak form \eqref{eq:weak_form} separately as the notation  in \eqref{eq:matrix_free_eval} might suggest at first glance. Instead, the gather and scatter operations as well as the interpolation and integration operations are performed only once for every cell, and they are combined with all the different quadrature point contributions (e.g. $\frac{\rho c}{\Delta t}\vec{I}$,  $\vec{D}_e$ and $\vec{L}_e$ in \eqref{eq:matrix_free_eval}). Thus, only one loop over all cells is required, increasing data locality.
	
	Additionally, we have adopted a
	concept from \cite{Kronbichler2022} which allows to
	load $\vec{T}_n$ only once from main memory during the evaluation of \eqref{eq:forward_euler}. For this purpose, we interleave cell operations and operations run on ranges of indices. Before an index $i$ is first used in the source vector $\vec{T}_n$, we clear the content of the destination vector for this index, $T_{n+1,i} \leftarrow 0$. Then, as an intermediate result, we assemble the right-hand side $T_{n+1,i} \leftarrow f_i(\vec{T}_n)$ via the fast operator evaluation. After all contributions to an index in this vector have been added, we run the update of the temperature vector  $T_{n+1,i} \leftarrow T_{n,i} + \Delta t  \tilde{C}_{ii} T_{n+1,i}$. This update is cache-efficient since  $T_{n,i}$ and $T_{n+1,i}$ are likely still in the cache from the operator evaluation.
	
	For the parallel, distributed computation we utilize MPI (Message Passing Interface). To see the highest possible throughput on the global level, the implementation in \texttt{deal.II} takes care of MPI communication and overlaps it with local computations.
	
	By basing our work on the high-performance implementation for operator evaluation in the \texttt{deal.II} library and actively contributing new features developed for our specific application, we are well-equipped for future extensions of our framework. The main feature that was added to the \texttt{deal.II} library in the context of this work is related to growing geometries by activating cells. The fact that some cells are inactive and do not carry any DoFs means that they must be skipped within the matrix-free evaluation framework. However, the interface between the active cells in the top-most layer and the inactive cells above them represents a boundary for the currently active domain and we want to evaluate the boundary conditions \eqref{eq:bc_radiation} and \eqref{eq:bc_evaporation} on these internal faces. These challenges are solved by the new class \texttt{ElementActivationAndDeactivationMatrixFree}, which allows to ignore non-active cells and perform integrals at faces shared by active and non-active cells. 
	
	\section{Results and discussion}
	
	All examples are run on our own compute cluster which consists of 52 compute nodes with a dual-socket Intel Xeon E5-2680 v3 CPU with 2 $\times$ 12 cores running at 2.5 $\si{\giga\hertz}$ and 8 DDR4 memory channels running at 2.13 $\si{\giga\hertz}$ (measured \textsc{STREAM} memory bandwidth of 82 $\si{\giga B\per\second}$). For this hardware the code is compiled with the AVX2 instruction-set extension. Importantly, we compared our model implementation on this hardware with a simple Laplace operator with constant coefficient implemented within the matrix-free \texttt{deal.II} framework. On one compute node, a full explicit time step of our model implementation reaches 44\% of the throughput of the Laplace operator. Note that our explicit operator performs significantly more computations (nonlinear material evaluation) and needs to load more data (material history data on quadrature points) than the Laplace operator. Thus, we can say that our implementation is already well-optimized and only further improvements in \texttt{deal.II} might give additional speedup in the future.

	\subsection{Bridge example}
	\begin{figure}
		\centering
		\includegraphics[width=.5\linewidth]{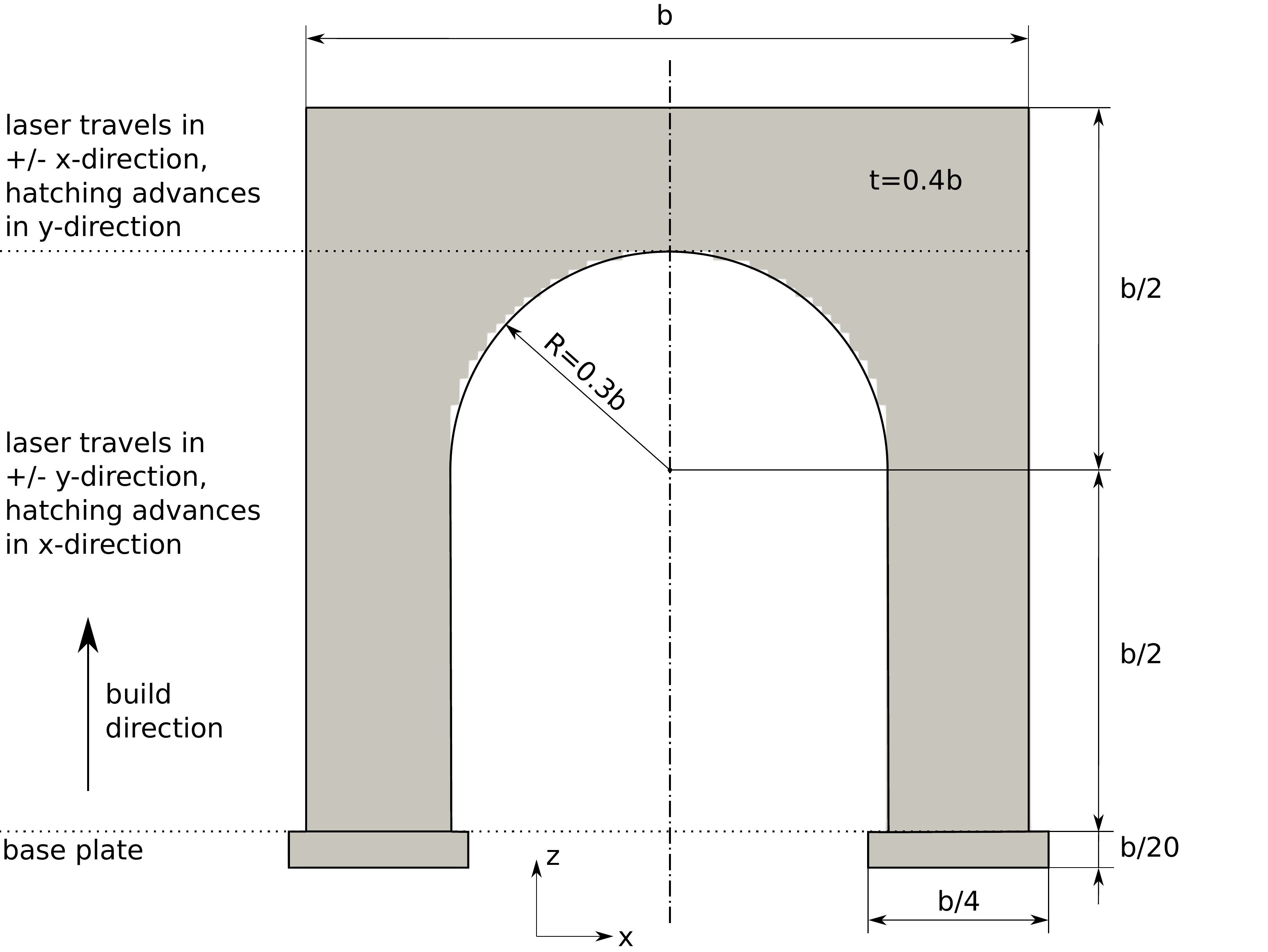}
		\caption{Bridge geometry: outline of ideal geometry design (solid line) and discretized voxel geometry (grey area). This geometry is investigated for different scalings controlled by a single parameter $b$.}
		\label{fig:bridge_geometry}
	\end{figure}
	As a first example we investigate a bridge geometry, schematically depicted in Figure \ref{fig:bridge_geometry}. We use a boundary-fitted voxel mesh that approximates the arc with a stair case profile. The geometry is parametrized by a single parameter $b$ to study different scalings of the problem. The coarsest cell size, which is equal to the voxel discretization size, is computed as $b/80$. The bottom of the base plate is kept at a fixed temperature $T_0$. The top surface (with positive $z$ normal vector) is subject to radiation \eqref{eq:bc_radiation} and evaporation \eqref{eq:bc_evaporation} boundary conditions. All other parts of the boundary are assumed to be thermally insulating, since they would be surrounded by powder (not modeled). Note that the small base plate section is intentionally reduced in size compared to a large, realistic base plate with dimensions in the decimeter scale, since we want its size to also scale with the parameter $b$. For the studies conducted in this work we found from our previous work \cite{Proell2021} that the effect of a large base plate on the global temperature response is negligible. Our framework is capable to include a large base plate, which is adaptively refined in the vicinity of the attached parts, should this become necessary in future validation examples.
	
	The scan pattern consists of serpentine tracks. The laser beam input parameters are given in Table \ref{tab:scan_params_bridges}. For the material behavior and radiation and evaporation boundary conditions, we choose values representative of the metals used in application, see Table \ref{tab:mat_params}. With these material parameters we obtain $\Delta t_\text{max} = \num{2.9e-4}\, \si{\second}$ as an estimate for the critical time step from condition \eqref{eq:explicit_time_step}. The actually used time step is $\Delta t = \num{2.0e-5}\si{\second}$ so that the laser beam travels half a cell (less than one laser beam radius of $R=\num{50}\si{\micro\meter}$) within one step. As mentioned earlier, for these material parameters and mesh sizes the stability limit \eqref{eq:explicit_time_step} turns out to be not restrictive compared to the accuracy requirement of the moving heat source \eqref{eq:moving_heat_source_condition}. After every layer, we simulate $\num{1} \si{\second}$ of interlayer cool down time as follows: the first 1000 time steps of the cool down phase are still simulated with the explicit time integration scheme of the active laser phase to capture the highly dynamic behavior. Afterwards, the time step is increased to $\Delta t = \num{2.0e-2}\si{\second}$ and the implicit time integration scheme is used to simulate the remaining $\num{0.98}\si{\second}$ of cool down time.

	\begin{table}
		\centering
		\caption{Scan parameters for bridge example.}
		\label{tab:scan_params_bridges}
		\begin{tabular}{llrl}
			\toprule
			Symbol & Property & Value & Unit\\
			\midrule
			$v_\text{scan}$ & Scan velocity & 1000 & $\si{\milli\metre\per\second}$\\
			$d_h$ & Hatch spacing & 80 & $\si{\micro\metre}$\\
			$R$ & Beam radius & 50 & $\si{\micro\metre}$\\
			$h_\text{powder}$ & Powder layer thickness & 40 & $\si{\micro\metre}$\\
			$t_\text{cool}$ & Cool down time & 1 & $\si{\second}$\\
			\bottomrule
		\end{tabular}
	\end{table}

	\begin{table}
		\centering
		\caption{Material parameters for bridge example. The parameters are representative for stainless steel and taken from our previous publication \cite{Meier2021a}.}
		\label{tab:mat_params}
		\begin{tabular}{llrl}
			\toprule
			Symbol & Property & Value & Unit\\
			\midrule
			$k_{ms}$ & Thermal conductivity in melt and solid phase & 20 & $\si{\watt\per\metre\per\kelvin}$ \\
			$k_{p}$ & Thermal conductivity in powder phase & 0.2 & $\si{\watt\per\metre\per\kelvin}$ \\
			$\rho$ & Density & 7430 & $\si{\kilogram\per\cubic\meter}$\\
			$c$ & Specific heat capacity & 965 & $\si{\joule\per\kilogram\per\kelvin}$ \\
			$T_s$ & Solidus temperature & 1500 & $\si{\kelvin}$\\
			$T_l$ & Liquidus temperature & 1900 & $\si{\kelvin}$\\
			$T_0, T_\infty$ & Initial and ambient temperature & 303 & $\si{\kelvin}$\\
			$\epsilon$ & Emissivity & 0.7 & --\\
			\midrule
			$T_v$ & Boiling temperature & 3000 & $\si{\kelvin}$ \\
			$C_P$ & Recoil pressure factor & 54 & $\si{\kilo\pascal}$ \\
			$C_T$ & Recoil pressure temperature factor & 50000 & $\si{\kelvin}$ \\
			$C_M$ & Heat loss temperature factor & 0.001 & $\si{\kelvin\square\second\per\square\meter}$ \\
		 $M$ &  Molar mass &  0.052 &  $\si{\kilogram\per\mole}$ \\
			$h_v$ & Latent heat of evaporation & 6.0 & $\si{\mega\joule\per\kilogram}$ \\
			$T_{h,0}$ & Enthalpy reference temperature & 663 & $\si{\kelvin}$ \\
			\bottomrule
		\end{tabular}
	\end{table}
	
	\begin{table}
		\centering
		\caption{Weak scaling study for bridge examples. Geometry and mesh information as well as performance results (only for the active laser phase).}
		\label{tab:weak_scaling_bridge}
		\begin{tabular}{rrrrrr|rrr}
			\toprule
			$b$ [mm] & build vol. [$\si{\cubic\milli\meter}$] & layers &  cores & $h_\text{coarse}$ [mm] & $n_\text{refine}$ & max. DoFs (per core) & time steps & wall time\\
			\midrule
			6.4 & 59 & 160 & 12 & 0.08 & 1 &  177k (14.8k) &956,504&  0.55 \si{\hour}\\
			12.8 & 469 & 320 & 48 & 0.16 & 2&  357k (7.44k) & 7,425,200& 2.7 \si{\hour}\\
			25.6 & 3,749 & 640 & 192  & 0.32 & 3 & 1,070k (5.57k) &59,187,712& 18.3 \si{\hour}\\
			51.2 & 29,991 & 1280 & 768 & 0.64 & 4& 3,850k (5.01k) &468,263,936& 128.9 \si{\hour}\\
			\bottomrule
		\end{tabular}
	\end{table}
	
	\begin{figure}
		\centering
		\includegraphics[width=\linewidth]{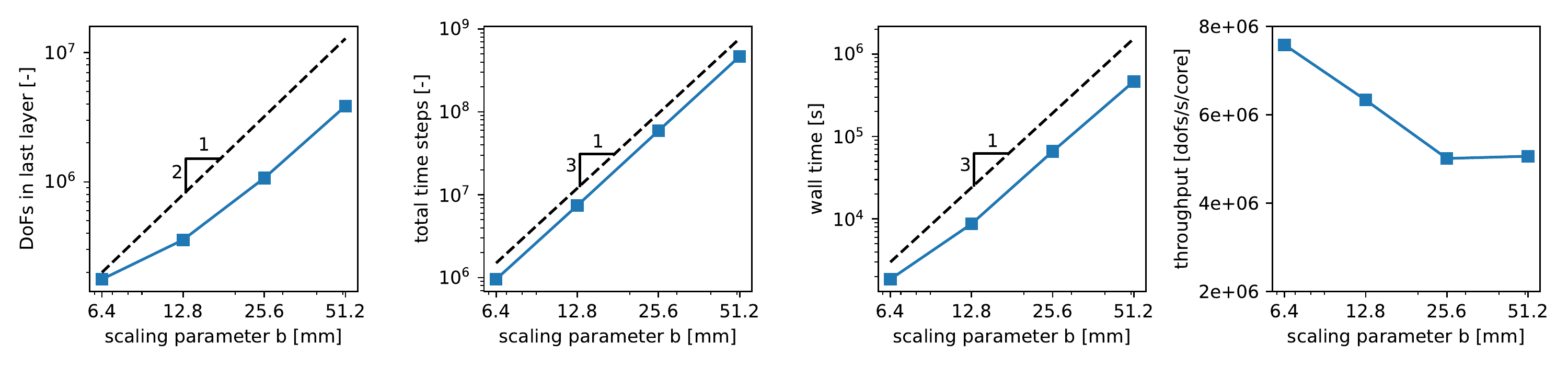}
		\caption{Scaling study for bridge example. The number of DoFs increases quadratically, while the number of time steps increases cubically. The wall time increases cubically because the number of CPU cores is increased such that the number of DoFs per core stays approximately constant. Therefore, the throughput in DoFs per second per core stays roughly constant as well. }
		\label{fig:weak_scaling_bridge}
	\end{figure}
	
	\begin{figure}
		\includegraphics[width=\linewidth]{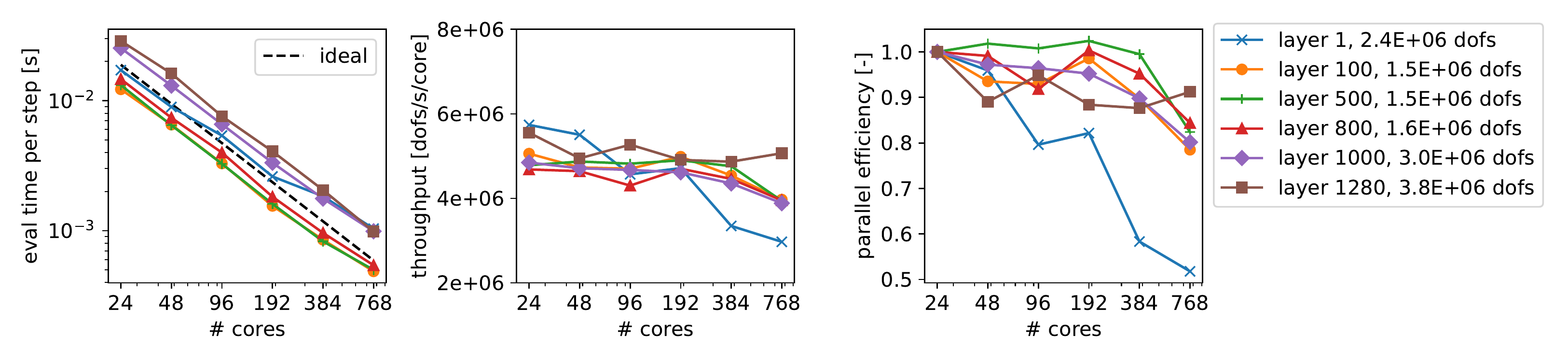}
		\caption{Strong scaling study for different layers of 1280 layer bridge example: evaluation time for a single time step (left), throughput (middle) and parallel efficiency (right). }
		\label{fig:strong_scaling_bridge}
	\end{figure}
	
	\begin{figure}
		\includegraphics[width=\linewidth]{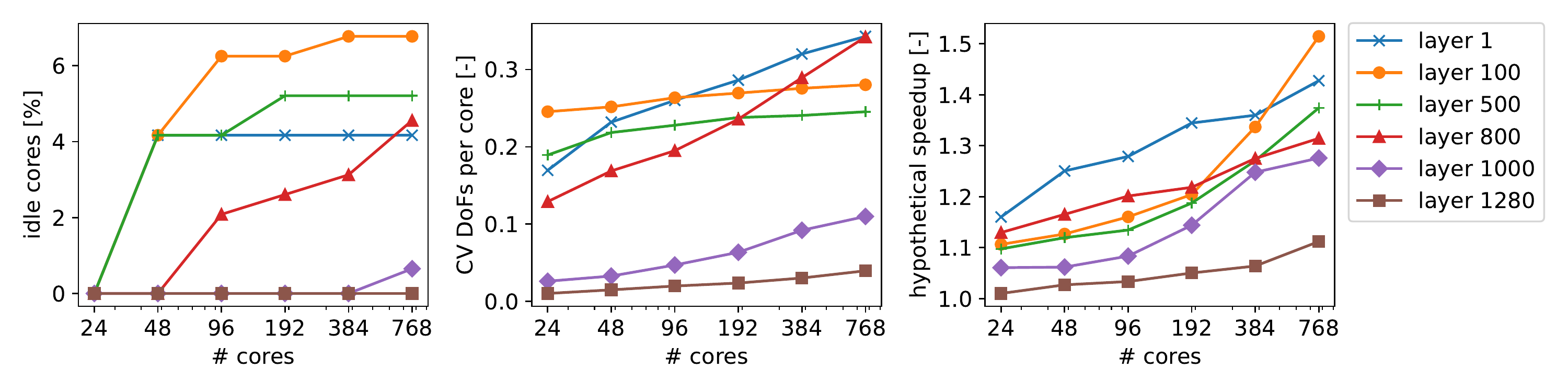}
		\caption{Imbalance in work across cores in different layers of 1280 layer bridge example: percentage of idle cores without any DoFs (left), coefficient of variation (CV) of the DoFs per core (middle) and hypothetical speedup, if the DoFs were distributed evenly among cores and additional communication overhead is neglected.}
		\label{fig:strong_scaling_bridge_imbalance}
	\end{figure}
	
	\begin{figure}
		\centering
		\includegraphics[width=\linewidth]{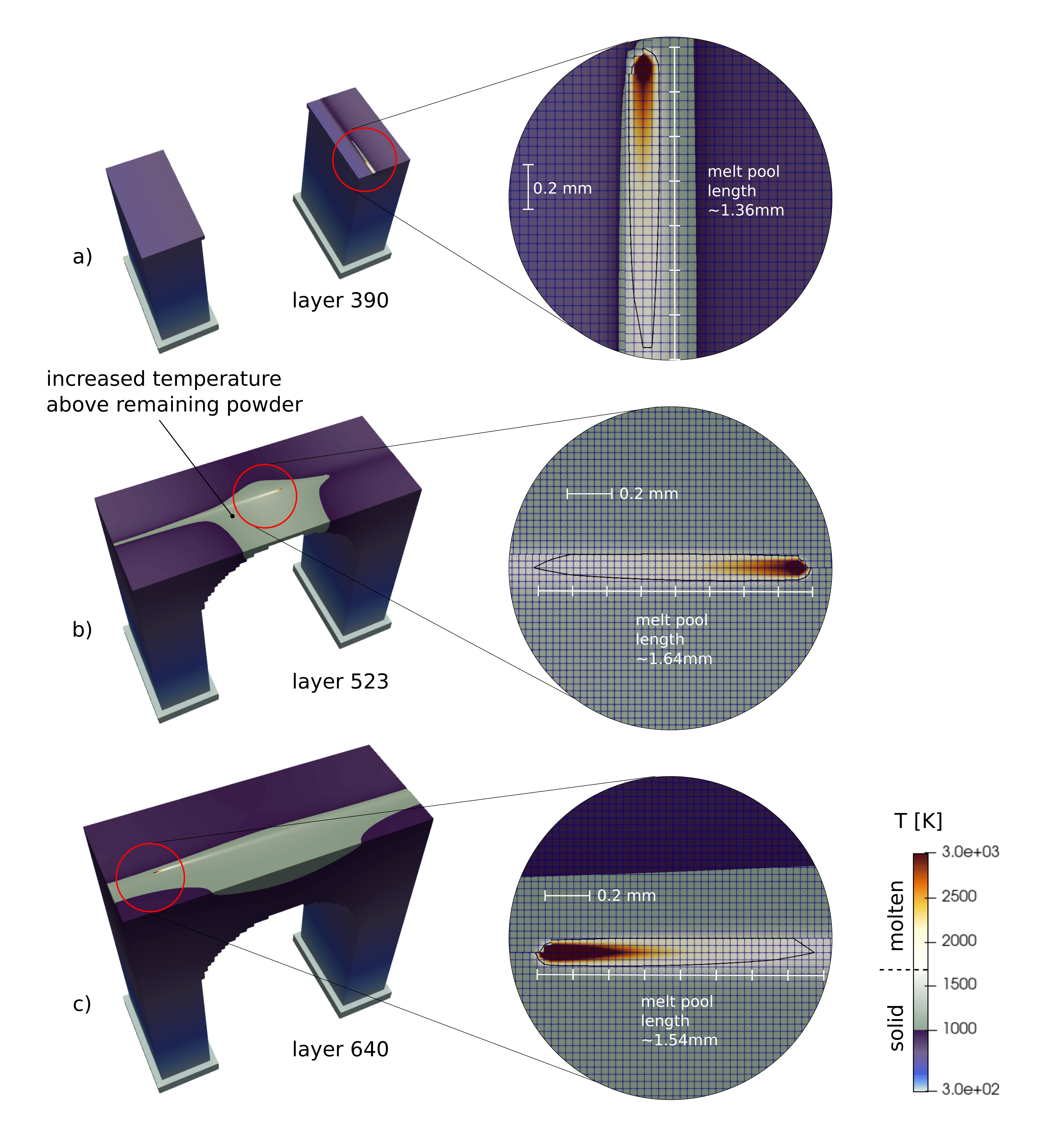}
		\caption{Temperature distribution in 640-layer variant of the bridge geometry at different stages of the build process. Regions with temperature above $T_m$ are defined as `molten' (surrounded by a black solid line), otherwise as `solid'. The heat affected zone (HAZ) with temperatures between 1000K and melt temperature is visualized in grey color. }
		\label{fig:bridge_temperatures}
	\end{figure}

	First, we perform a type of weak-scaling study, where we increase the dimension parameter $b$ as indicated in Table \ref{tab:weak_scaling_bridge} and at the same time increase the computational resources. The scaling of the mentioned quantities is also visualized in Figure \ref{fig:weak_scaling_bridge}.  The scalability of the implementation in the \texttt{deal.II} library has already been demonstrated \cite{arndt2020exadg, Kronbichler2018}. The goal of this type of scaling study is to illustrate the various scaling effects that occur in a part-scale PBFAM simulation.  Since we repeatedly increase the domain size by a factor of 2 in each dimension (via parameter $b$), the build volume increases cubically, i.e., by a factor of 8. Remember that -- compared to many other works -- we do not artificially scale up the heat source (nor the layer thickness) in this contribution. Therefore, the length of the complete laser track and consequently the number of time steps scales directly with the build volume, i.e., it increases by factor 8 as well (Figure~\ref{fig:weak_scaling_bridge}, panel 2). Notice that, as the domain grows larger and more refinement levels are necessary, the number of cells and total DoFs only increases by a factor of approximately 4 per scaling step (Figure~\ref{fig:weak_scaling_bridge}, panel 1). This is a consequence of AMR: the number of top-most layers with the highest refinement is constant. When the geometry is scaled up, a relevant increase of DoFs only happens in these top-most layers in the $x$- and $y$-directions
	
	To understand how our framework behaves for growing domain sizes, we scale the computational resources by a factor of 4, i.e., the expected scaling of the number of DoFs in the later layers. The reasoning behind this choice is the good parallel scalability of the spatially distributed single-step evaluation: by scaling the computational resources by the same factor as the number of DoFs, we keep the work per process approximately constant in the later scaling steps.  Therefore, assuming weak scalability, we expect the wall time to grow proportional to the number of time steps, i.e., by a factor 8 per scaling step, as observed in Figure~\ref{fig:weak_scaling_bridge}, panel 3.  Perfect weak scaling is observed for the last scaling step where the throughput, defined as
    \begin{align}
    \text{throughput} = \frac{\text{number of DoFs}}{\text{eval time per step} \times \text{number of cores}},
    \end{align}
    stays constant (Figure~\ref{fig:weak_scaling_bridge}, panel 4).
	Note that we cannot simply scale up the computational resources by an additional factor 8 to counteract the increased number of time steps, since the work is not parallelized in time.

	As a second study, we investigate the strong scaling capabilities of the framework. To this end, the largest geometry of the first study (with $b=51.2 \si{\milli\meter}$) is simulated with a varying number of CPU cores $n_\text{core} \in \lbrace 24, 48, 96, 192, 384, 768\rbrace$. Since we are now interested in the scalability over different layers, we only simulate 1000 steps of the active laser phase per layer. The results are transferable to the whole layer because layers are activated in full at the beginning of a new layer and all cells are immediately active (though still in powder state).
	
	The resulting average evaluation time for a single step in different layers is shown in Figure \ref{fig:strong_scaling_bridge}. The strong scaling is close to the ideal behavior in the higher layers  and, in the lower to medium layers, we are approaching the scaling limit of $\num{1e-4}\,\si{\second}$ reported in \cite{Kronbichler2018}. In the first layer, there is not enough work for the assigned cores such that an increased number of cores does not result in an equivalent speedup. This is further illustrated by the total throughput (measured in DoFs per second per core per time step) and the parallel efficiency $\eta$, defined here as 
	\begin{align}
	\label{eq:parellel_efficiency}
	\eta = \frac{T_\text{ref}N_\text{ref}}{T_\text{scaled}N_\text{scaled}},
	\end{align}
	where $T_\text{ref}$ and $N_\text{ref}$ are the wall time and resources used for a reference run (in this case, a run on one compute node with 24 CPUs). $T_\text{scaled}$ is the wall time for a run with $N_\text{scaled} = sN_\text{ref}$ ($s$ times more) resources.
	The parallel efficiency $\eta$ is a measure for the efficient use of resources, where a value of 1 means the additional resources manifest in a perfect speedup. In Figure \ref{fig:strong_scaling_bridge}, the parallel efficiency is mostly close to 1. It drops to around 50\% for the first layer when run with the largest number of cores. However, in the last layer 1280, the parallel efficiency stays at around 90 \% even for the highest core count which justifies the use of these computational resources. Note that we sometimes see a parallel efficiency slightly greater than one, e.g. in layer 500. This happens since the work per process and especially the number of ghosted cells varies with the layer number and the number of processes which can lead to a case where communication overhead is slightly worse for smaller core counts.
	
	More detailed insights into the imbalance of the DoF distribution among cores is shown in Figure \ref{fig:strong_scaling_bridge_imbalance}. Only a small fraction of cores is idle, i.e, has no DoFs at all. This fraction increases with larger core counts and decreases in the higher layers. As another metric, the coefficient of variation (CV) of the DoFs assigned to a core is defined as the ratio between standard deviation and mean of that same quantity. The CV reveals a strong imbalance in layer 800, which once again shows that the quality of the partitioning is layer-dependent. We can estimate an upper bound for the hypothetical speedup obtained by a better distribution, if we divide the maximum number of DoFs per core by the mean. This yields the speedup factor for an even distribution of DoFs among cores when additional communication overhead is neglected. The hypothetical speedup is at most 1.5 (layer 100 in Figure \ref{fig:strong_scaling_bridge_imbalance}), although it can never be fully realized and the estimate is very optimistic. Since this hypothetical gain is lower in most layers, we did not yet work on a more optimal parallel distribution in this paper. A simple weighting of active cells by a factor of 10 or 100 (compared to inactive cells) combined with a parallel redistribution when a new layer is activated did not produce a notable speedup which is in line with results reported in \cite{Neiva2019} for an adaptive mesh similar to the one investigated here.
	
	\begin{table}
		\centering
		\caption{Single-step solution time and relative speedup for different time stepping schemes and vectorization levels in the active laser phase. To get a direct comparison of evaluation costs, a linear problem is solved such that the implicit$^\ast$ scheme only performs one nonlinear iteration.}
		\label{tab:rel_speedup}
		\begin{tabular}{r|ccc|ccc}
			\toprule
			& \multicolumn{3}{c}{\num{26e3} DoFs/core} & \multicolumn{3}{c}{\num{6.5e3} DoFs/core}\\
			& no vectorization& & 4-wide SIMD &  no vectorization& & 4-wide SIMD\\
			\midrule
			implicit$^\ast$ & 1.07 s & $\xrightarrow{\times 1.79}$ & 0.598 s& 0.323s & $\xrightarrow{\times 1.77}$ & 0.182s \\
			& $\downarrow$ \scriptsize $\times 99.1$  & &  $\downarrow$ \scriptsize $\times 114$ &$\downarrow$ \scriptsize $\times 89.2$  & &  $\downarrow$ \scriptsize $\times 143$ \\
			explicit & 0.0108 s & $\xrightarrow{\times 2.07}$ & 0.00524 s & 0.00362s & $\xrightarrow{\times 2.85}$ & 0.00127s\\
			\bottomrule
		\end{tabular}
	\end{table}

    In order to analyze the impact of different implementation aspects, we present relative speedup data in Table~\ref{tab:rel_speedup}. This data was obtained from running the active laser phase in the last layer of the 1280 layer bridge example. For a fair comparison, we deliberately do not include a comparison with less optimized code and all cases use the matrix-free, highly-optimized code infrastructure from \texttt{deal.II}. Also, we choose the parameters such that the implicit system is linear and solved within a single nonlinear iteration to get a better comparison of the inherent cost associated with an implicit linear solve step. Since the actual problem is nonlinear, in practice, the evaluation costs for an implicit scheme are even higher when a few nonlinear iterations are required. As Table~\ref{tab:rel_speedup} reveals, an explicit step is around 100 times cheaper than an implicit step. However, it should be emphasized that our implicit scheme together with the preconditioner for the linear solver has not been optimized to the same degree as the explicit scheme since it does not present a bottleneck when only using it for the cool down phase. The speedup obtained from vectorization is more pronounced for the explicit than the implicit scheme since the operator evaluation is fully vectorized while the implicit linear solver also contains unvectorized code. Notably, for the explicit scheme the benefit of vectorization is higher at a lower load per CPU core because the problem is small enough to fit in the cache which in turn makes SIMD parallelism more impactful. The results further illustrate how the implementation approach scales well also for low loads per CPU core.

	To conclude this section, we show the temperature distribution on the bridge geometry with 640 layers in Figure \ref{fig:bridge_temperatures}. In the beginning of the AM process, when the two legs are still separated, the region of high temperatures in the solid material (indicated in grey color) is localized around the melt pool (Figure \ref{fig:bridge_temperatures}a)). Once the legs join up into a continuous layer the heat affected zone (HAZ) -- with temperatures between 1000K and melt temperature visualized in grey color -- stretches over the strongly overhanging middle region on top of a powder domain, which is thermally insulating in good approximation (Figure \ref{fig:bridge_temperatures}b)). Note the complex and asymmetric shape of this region, which can only be captured by a scan-resolved simulation as performed in this work. In the future, the model can be extended to predict the influence of this overheated region on the microstructure evolution and, ultimately, residual stress formation. In the last layer (see Figure \ref{fig:bridge_temperatures}c)) the enlarged high-temperature region still persists. This is a result of the parameters chosen in this example, especially the cool down time between layers, which is 1 \si{second} in this example. As a result, the initial temperature when the scanning of a new layer begins, increases with increasing number of layers.
	
	To further motivate the utility of the model in the present state let us mention a few more highly relevant and AM-specific effects that can be studied with it. As seen in this example, the melt pool length can be easily extracted from the temperature results (or, even be calculated while running the simulation) which allows a prediction of balling due to the Plateau-Rayleigh instability. An analysis of the peak temperatures would allow us to predict zones of excessive evaporation and gas-bubble-induced porosity. Residual porosity due to lack-of-fusion can be directly determined from the consolidation history (as shown in the next example). The presented model is still a part-scale model and all of the mentioned effects are captured in a qualitative manner. For a detailed quantitative analysis one needs to resort to mesoscale models.

	\subsection{Cantilever benchmark}
	
	\begin{figure}
		\centering
		\includegraphics[width=.7\linewidth]{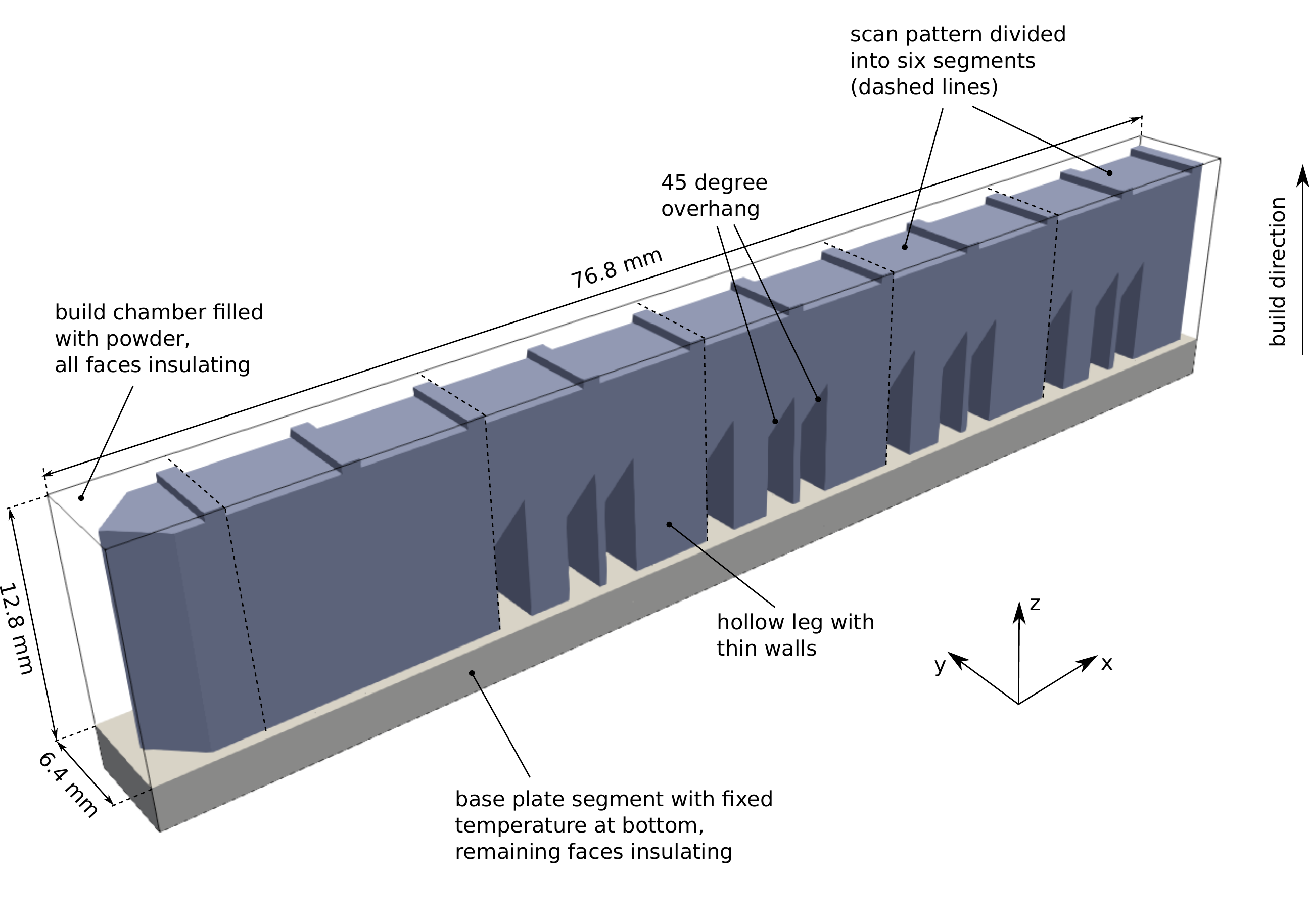}
		\caption{Overview of NIST AM Benchmark 2022 cantilever geometry and special features. Detailed information about the dimensions, geometry and scan strategy can be found in \cite{AMbench2022}.}
		\label{fig:cantilever_setup}
	\end{figure}%
	
	\begin{table}
		\centering
		\caption{Scan parameters for cantilever example.}
		\label{tab:scan_params_cantilever}
		\begin{tabular}{llrl}
			\toprule
			Symbol & Property & Value & Unit\\
			\midrule
			$v_\text{scan}$ & Scan velocity & 960 & $\si{\milli\metre\per\second}$\\
			$d_h$ & Hatch spacing & 110 & $\si{\micro\metre}$\\
			$R$ & Beam radius & 60 & $\si{\micro\metre}$\\
			$h_\text{powder}$ & Powder layer thickness & 40 & $\si{\micro\metre}$\\
			$t_\text{cool}$ & Cool down time & 1 & $\si{\second}$\\
			\bottomrule
		\end{tabular}
	\end{table}%
	
	As a second numerical example, we investigate the cantilever structure shown in Figure~\ref{fig:cantilever_setup}, which was designed for the NIST AM Benchmark series 2022 \cite{AMbench2022}.  The purpose of this example is not yet to validate the model against experimental measurements. Rather we want to demonstrate the capabilities of the framework on realistic geometries. Since the geometry is more complex than in the previous example, we use a build chamber mesh of dimensions $76.8\times 6.4 \times 12.8\,\si{\cubic\milli\metre}$ which also discretizes the remaining powder. The path of the laser beam is used as an input for the active laser phase and the tracks are scanned into a box that encloses the desired geometry such that a small buffer of remaining powder lies around the final part. Such a tightly fitting build chamber mesh is deemed acceptable due to the negligible powder conductivity. A base plate section of $2.56 \si{\milli\meter}$ thickness is added below the build chamber. Its bottom face is kept fixed at the initial temperature $T_0$.

	The active scan phase in every layer is followed by a cool down phase of $1 \si{\second}$, which uses the same time step sizes as described in the last section. After simulating all 312 layers, the built geometry is implicitly defined by the solid phase fraction at each quadrature point according to equation \eqref{eq:phase_fractions}. The scan parameters are given in Table~\ref{tab:scan_params_cantilever} while the same material parameters as for the bridge examples are reused, see Table~\ref{tab:mat_params}.

	\begin{figure}
		\centering
		\includegraphics[width=\linewidth]{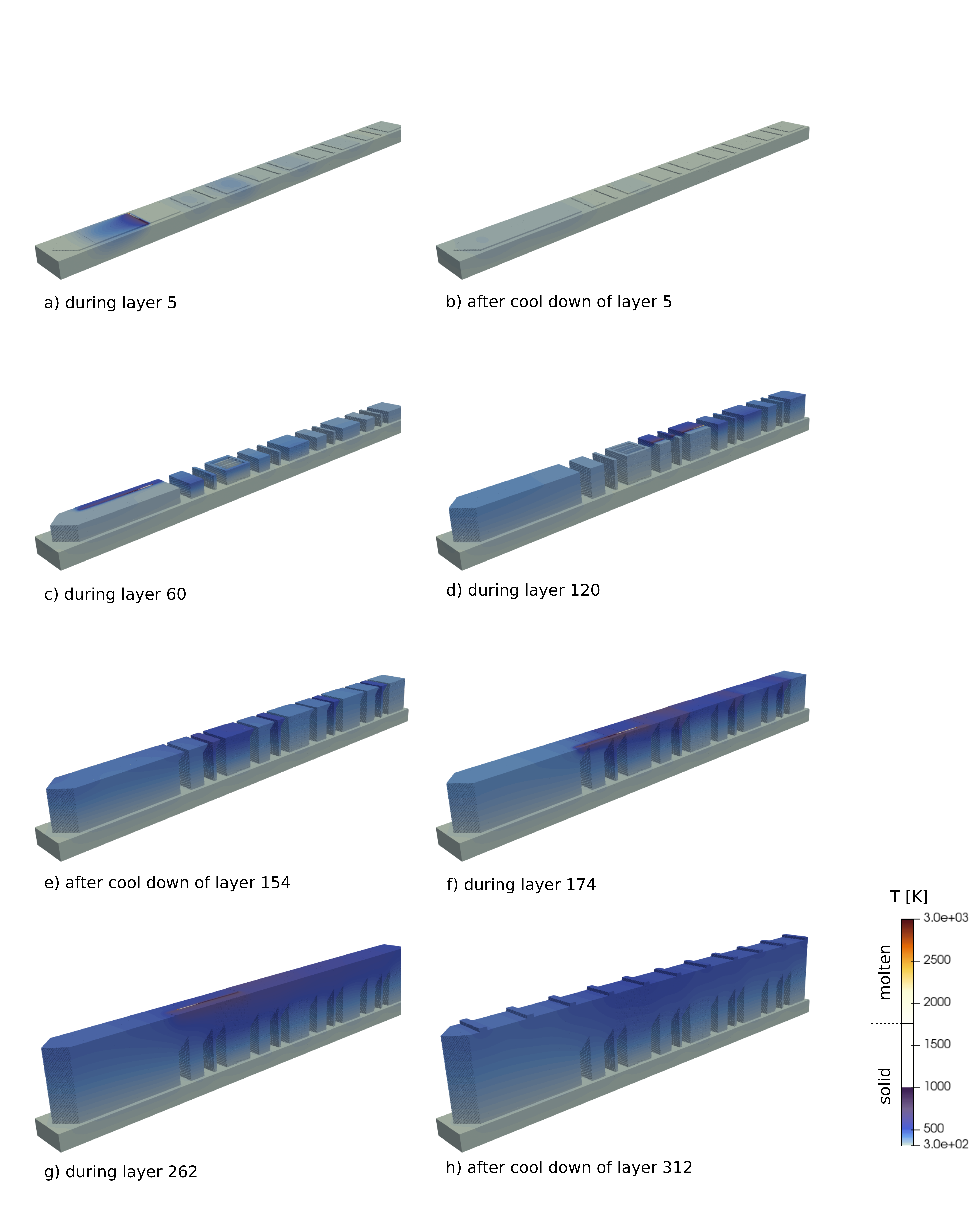}
		\caption{Temperature distribution in cantilever at various stages in the build process. Regions with temperature above $T_m$ are defined as `molten', otherwise as `solid'. A video of the complete process is attached as supplementary material to this article.}
		\label{fig:cantilever_temperature}
	\end{figure}
	
	\begin{figure}
		\centering
		\includegraphics[width=\linewidth]{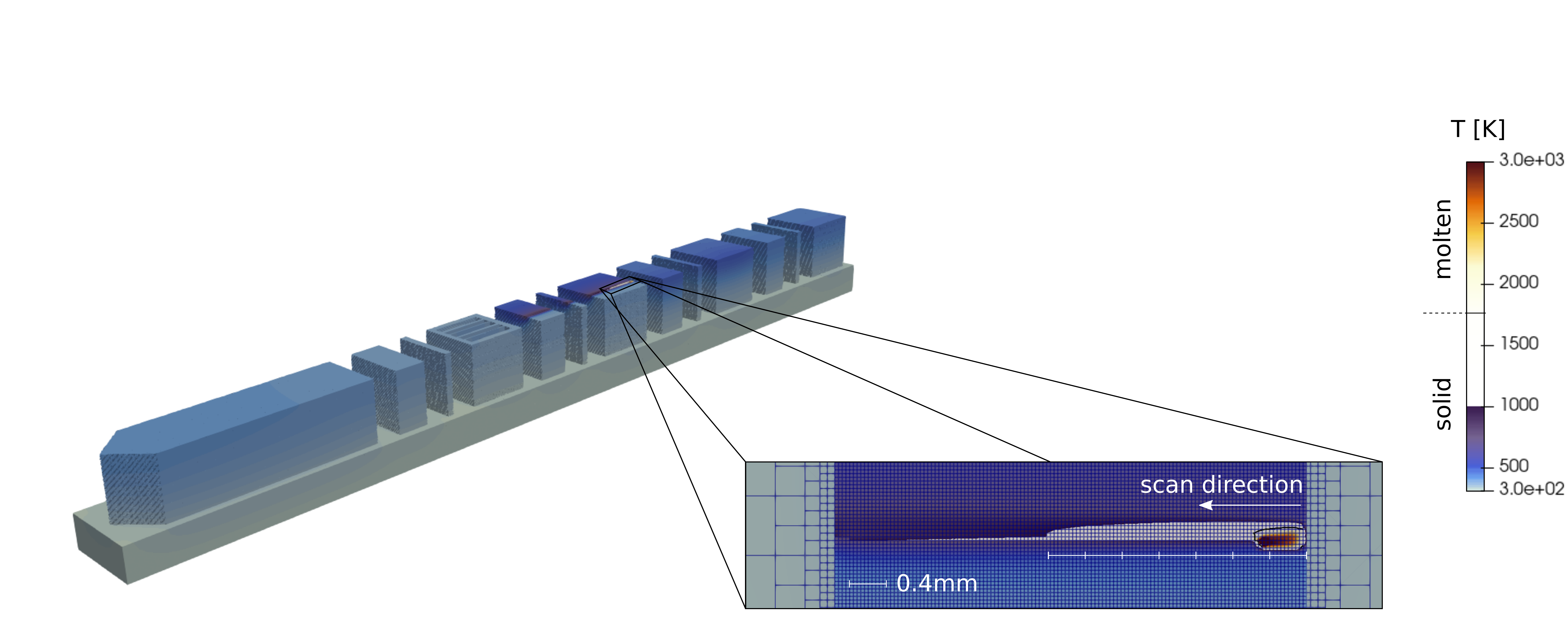}
		\caption{Detailed view of the melt pool (indicated by solid black line) at a turning point and start of a new track segment during processing of layer 120. Regions with temperature above $T_m$ are defined as `molten', otherwise as `solid'.}
		\label{fig:cantilever_melt_pool_detail_120}
	\end{figure}
	
	\begin{figure}
		\centering
		\includegraphics[width=\linewidth]{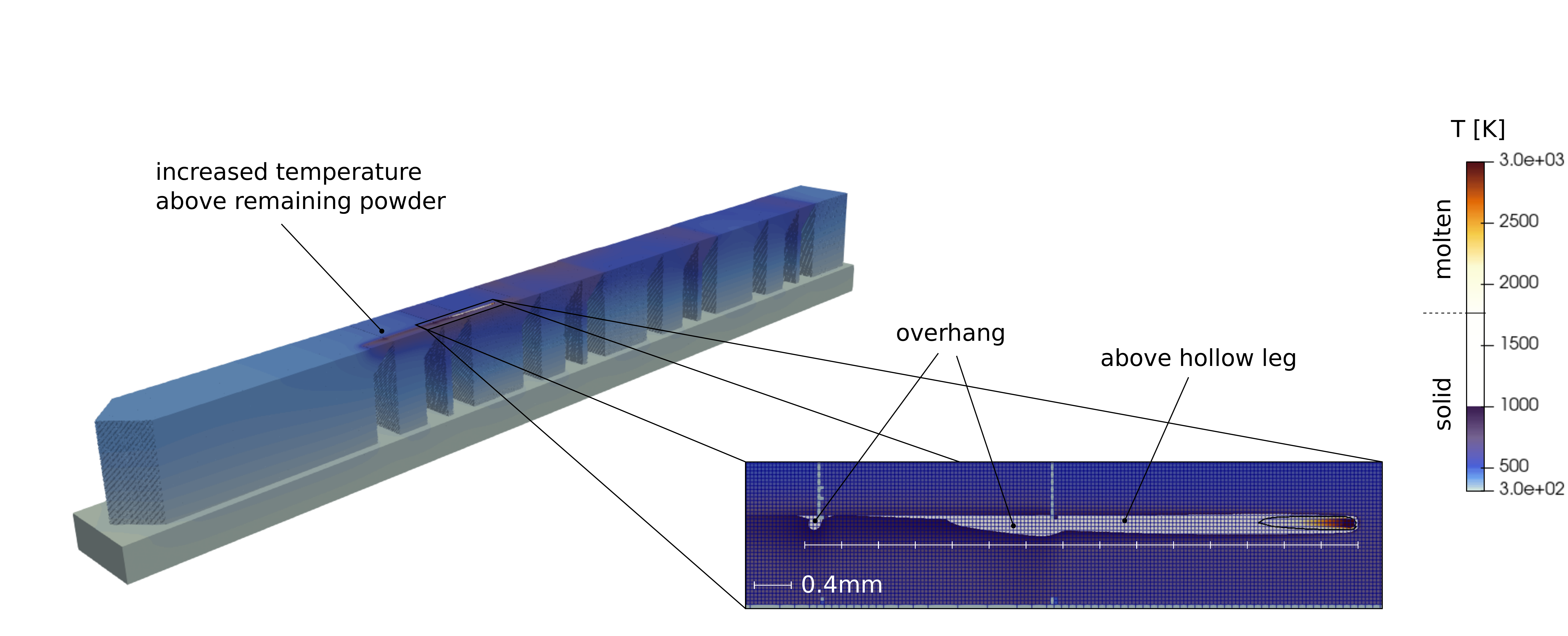}
		\caption{Detailed view of the melt pool (indicated by solid black line) during processing of layer 174. In this layer the separate legs join up into a continuous layer. Regions with temperature above $T_m$ are defined as `molten', otherwise as `solid'.}
		\label{fig:cantilever_melt_pool_detail_174}
	\end{figure}

	\begin{figure}
		\centering
		\includegraphics[width=\linewidth]{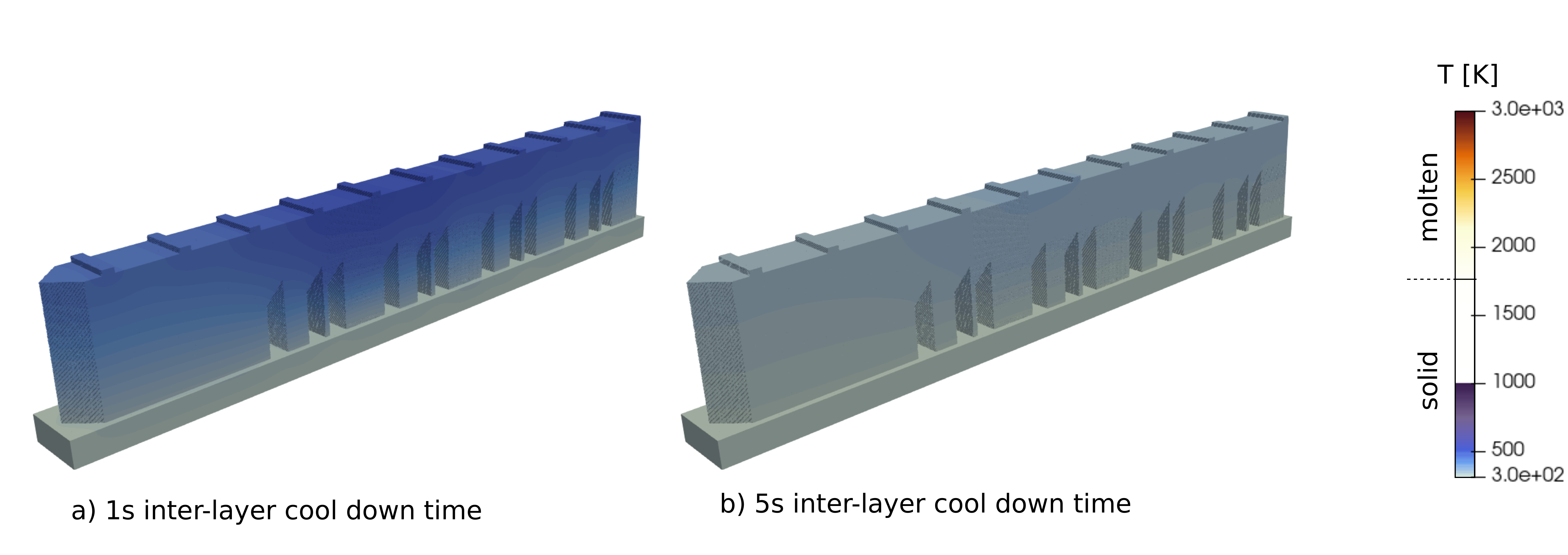}
		\caption{Residual heat after completed build process a) for  $1\si{\second}$ cool down time and b) for 5 $\si{\second}$ cool down time.}
		\label{fig:cantilever_residual_heat_1s_vs_5s}
	\end{figure}
	
	\begin{figure}
		\centering
		\includegraphics[width=\linewidth]{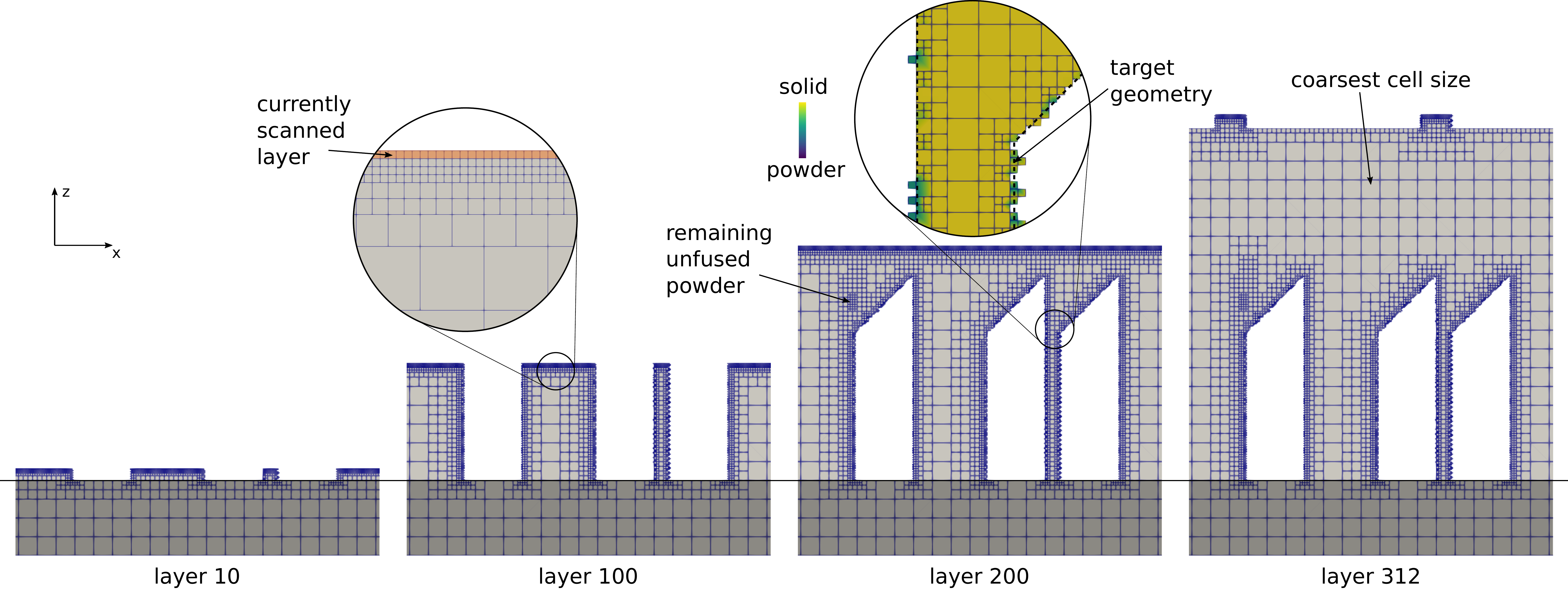}
		\caption{Detailed view of the part buildup and adaptive mesh in the symmetry ($xz$-)plane of the cantilever. The finest mesh resolution is only kept when necessary to capture the final part shape. Any cell with more than 50 \% solid fraction is visualized.  The baseplate is overlaid in dark grey.}
		\label{fig:cantilever_mesh_evolution}
	\end{figure}
	
	\begin{figure}
		\centering
		\includegraphics[width=\linewidth]{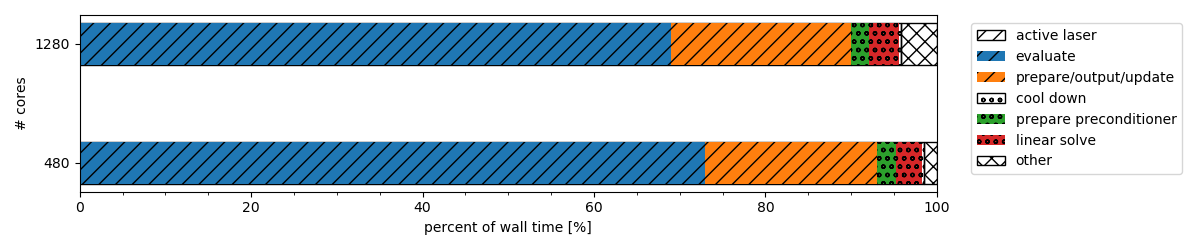}
		\caption{Percentage of wall time spent in the different parts of the algorithm for the 1248 core simulation compared to the 480 core simulation. Timings averaged over all 312 layers.}
		\label{fig:cantilever_timing}
	\end{figure}

	\begin{figure}
		\centering
		\includegraphics[width=\linewidth]{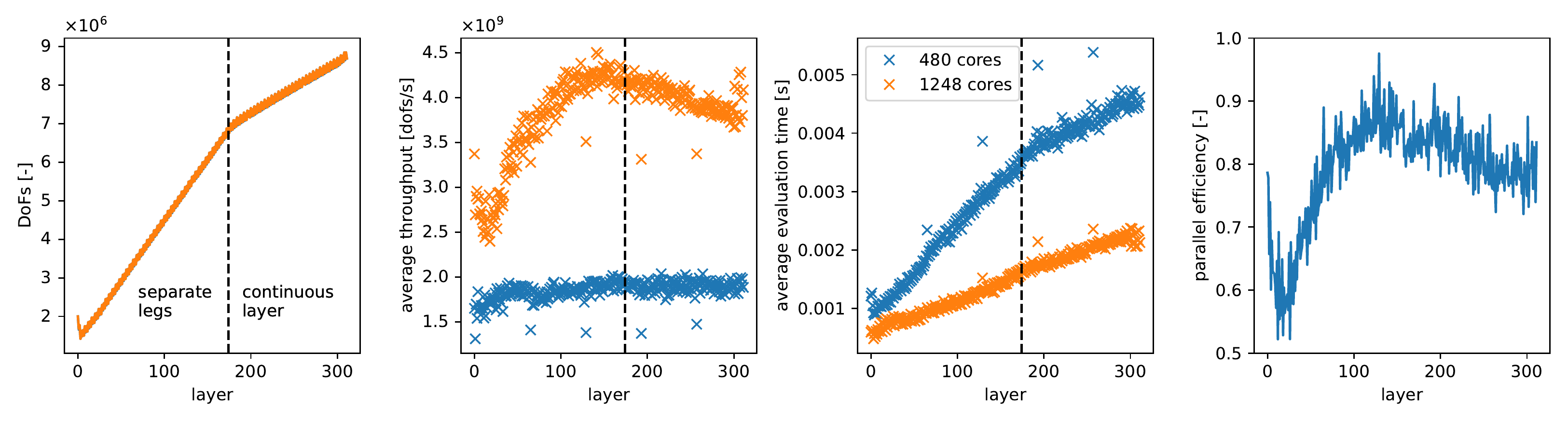}
		\caption{Number of DoFs, throughput, average evaluation time per step and parallel efficiency over all 312 layers of the cantilever example for the 1248 core simulation compared to the 480 core simulation. Layer 174, where the separate legs join up into a continuous layer, is indicated by a dashed line. The plotted data only includes the active laser phase solved with the explicit time stepping scheme.}
		\label{fig:cantilever_stats}
	\end{figure}

	The temperature distribution in the part at difference points in time of the process is shown in Figure~\ref{fig:cantilever_temperature}. A video of the complete process may also be found in the supplementary material of this article. In the first layers the temperature quickly drops to the initial temperature $303 \si{\kelvin}$ after the laser passed, see Figure~\ref{fig:cantilever_temperature}a-b). This fact can be explained by the small total heat capacity of the consolidated material and the small distance to the base plate with prescribed temperature at its bottom face. As more and more layers are processed, the residual temperature steadily rises but the HAZ with $T > 1000 \si{\kelvin}$ (indicated in white color) is always limited to the direct vicinity of the melt pool (Figure~\ref{fig:cantilever_temperature}c-d)). High temperature gradients are thus also limited to this area which justifies the use of a refined mesh only in these areas. A detailed view of the melt pool in layer 120 is shown in Figure~\ref{fig:cantilever_melt_pool_detail_120}. Figure~\ref{fig:cantilever_temperature}e) illustrates the different cooling rates resulting from the different geometrical features: the thin legs are at an elevated temperature compared to the thicker leg due to the smaller thermal conductivity (and heat flux concentrations) in these regions. An exception is the thick but internally hollow leg which exhibits an equally poor heat conduction to the base plate as the thin legs.
	
	In layer 174 the initially separate legs join up into a continuous layer (Figure~\ref{fig:cantilever_temperature}f)). The melt pool and the HAZ surrounding it are elongated when the beam travels over the overhanging regions and the hollow leg. A detailed view of the melt pool is shown in Figure~\ref{fig:cantilever_melt_pool_detail_174}. For the depicted point in time, the laser beam travels across the hollow leg and previously across an overhang region which leads to an elongated melt pool and an enlarged HAZ due to the decreased conductivity to the base plate. This effect persists in higher layers whenever the laser beam is moving across the hollow leg (Figure~\ref{fig:cantilever_temperature}g)). Note that such geometrical influences cannot be predicted with layer-wise part-scale models or melt pool models but only with scan-resolved, true part-scale models. The residual heat after cool down of the final part is shown in Figure \ref{fig:cantilever_temperature}h). In Figure~\ref{fig:cantilever_residual_heat_1s_vs_5s} we compare the final temperature distribution when  the (previously used) cool down time of $\num{1}\,\si{\second}$ or an (alternative) cool down time of $\num{5}\,\si{\second}$ is used after every layer. The cool down time has a strong influence on the temperature level after cool down. A systematic investigation of the impact of cool down time is thus possible with the approach presented in this work since the real scan and cool down times are used consistently in this model (in contrast to layer-wise simulation approaches, where heating and cooling times are often calibration parameters).
	
	To illustrate the adaptive meshing strategy, Figure~\ref{fig:cantilever_mesh_evolution} shows the growing part geometry on a slice. The coarse mesh uses a cell size $h_\text{coarse} = \num{640}\, \si{\micro\meter}$, i.e., four refinement levels are necessary to reach a cell size of $h_\text{powder}$. The number of DoFs grows linearly with the number of layers, see Figure \ref{fig:cantilever_stats}, with a visible kink once the initially separate legs of the cantilever join up into a continuous layer in layer 174. Since the interface surface between solid and powder is larger as long as the legs are separated and consequently more refined cells are necessary to capture this interface, the number of DoFs grows more quickly before layer 174 than afterwards. Figure \ref{fig:cantilever_mesh_evolution} also shows how well the build chamber mesh in combination with the adaptive mesh strategy can capture the final part geometry: the detail view in layer 200 overlays the target geometry outline over the built geometry in the overhang region. The solidified part shape agrees very well with this target geometry. Only in some places a small amount of partially molten powder sticks to the surface. Note that we can also capture an area where lack of fusion occurred and a refined mesh remains necessary. This illustrates a further aspect relevant for AM applications that can be predicted on the part-scale with the help of this model.
	
	The example was run on the same hardware as the previous bridge example. The simulation of the process for the total build volume of $3770\, \si{\cubic\milli\metre}$ requires around 44 million time steps and a maximum of  8.7 million DoFs. With 480 CPU cores, the total simulation time was 51.9 $\si{\hour}$. As an extension to the strong scaling study performed in the last section we increased the number of cores to 1248 (all cores available on the test machine), giving a total simulation time of 23.7 $\si{\hour}$ (which is an overall parallel efficiency of 84\% compared to the previous run). In both cases, most of the time is spent in the active laser phase, with the cool down phase ($\num{1}\si{s}$ of simulated time per layer) only taking around 5\% of the total wall time, see Figure \ref{fig:cantilever_timing}. The implementation of the cool down phase has not been optimized as much as the active laser phase in this contribution since we would not see a significant overall speedup in our numerical examples. Note that in another recent and performant model for scan-resolved part-scale analysis \cite{Olleak2022}, the authors simulate only four sets of four successive layers of a similar cantilever geometry. No exact timings are given. To the best of the authors' knowledge, no other scan-resolved model has been published so far which simulates, in practice, the build process on the scale of a real part.  For numerical examples with around 2 million DoFs other authors report single step solution times in the range of a few seconds in \cite{Neiva2019} or a few hundred milliseconds in \cite{Dugast2021}. By comparison, our presented approach is several orders of magnitude faster.

	The average throughput of the evaluation in terms of DoFs per second and the average time for one time step are also shown in Figure \ref{fig:cantilever_stats}. The throughput in the first few layers for the high core count indicates that we initially underutilize the assigned computational resources. Indeed, when examining the parallel efficiency for every layer separately it becomes clear that the core count and distribution of the problem could be improved in the first layers. As already mentioned, in the future dynamic resource allocation could be used so that shared computational resources are only used when necessary. In the later layers we reach a very good parallel efficiency of 80--90 \%, which still justifies the use of the increased computational resources.
	
	Note that the throughput and evaluation time in Figure \ref{fig:cantilever_stats} show outliers which occur exactly every 64 layers. In these layers, processes need a comparatively large number of ghosted information from other processes. This behavior can be linked back to the way \texttt{deal.II} distributes the cells and DoFs among processes \cite{Burstedde2011}, which might lead to non-contiguous subdomains and a non-uniform distribution of expensive hanging nodes \cite{Munch2022}. Again, fine tuning might give a speedup in the problematic configuration but is not further investigated in this work since on a global view it would only give a negligible speedup.

	\section{Conclusion and outlook}
	
	A high-performance approach for the simulation of part-scale laser powder bed fusion additive manufacturing (LPBFAM) with a resolved scan track has been proposed. The physics-based model includes phase-dependent material parameters and consistent boundary conditions. The dynamic heat equation is discretized with an explicit time stepping scheme which has a smaller computational cost per time step and better parallel scalability compared to implicit schemes. The stability limit inherent to explicit schemes is found to be less restrictive than the restriction imposed by the moving heat source (which, in the scan-resolved regime, should not travel further than its own radius within one step).
	
	We studied numerical aspects on basis of weak and strong parallel scaling tests. The implementation shows excellent scalability on a moderately-sized distributed compute cluster. Due to the explicit time stepping scheme and the high-performance implementation the time to solution for application-relevant problems is superior to other implementations in the literature that try to solve this problem. Notably, we achieve wall clock times per time step of a few milliseconds which is several order of magnitudes  lower than the timings reported in other implementations in the literature.

	Although we were able to reduce the cost of a single time step significantly within this work, scan-resolved simulation of LPBFAM parts on the scale of several decimeters or more most likely remains unrealistic due to the excessive number of time steps to be solved. Therefore, in future work the advances in this contribution could be combined with techniques that try to tackle the temporal scale, such as parallel-in-time methods \cite{Hodge2020, Moran2021} or space-time formulations \cite{kopp2022space}.

	With the presented adaptive mesh refinement strategy, using either a boundary-fitted or build chamber mesh, we are able to simulate general problem settings of LPBFAM as demonstrated on two realistic AM geometries. Notably, we performed the first full scan-resolved simulation of the NIST AM Benchmark cantilever in just below one day. Since the framework does not make any strong physical assumptions that require detailed calibration (such as layer-wise heat source models would), the obtained results already show interesting physical effects that are relevant for designers. A validation with real material data against measurements is a next step.
	The proposed thermal simulation model can serve as a basis for microstructure predictions on the part-scale, but also to study the influence of scan pattern and part geometry on melt pool shape and temperature, which are important indicators for process defects. These opportunities have been indicated throughout the discussion of the results.
	
	A natural extension of the current framework will deal with the thermo-mechanical problem. The groundwork has been laid in our contribution \cite{Proell2021} and needs to be incorporated into the high-performance framework presented in this work. Matrix-free implementations with efficient solution strategies exist for the solid mechanics problem \cite{Brown2022,Davydov2020}. They will likely require application-specific adaptations to complement the high-performance implementation of the thermal problem presented in this work.

	Although the current implementation can be said to be optimized when compared to a benchmark, a few performance-related topics for future investigation remain unresolved. In this work, we only looked at performance on CPUs. Due to the increasing popularity and availability of powerful GPUs, a compliant implementation might make the methodology available to a wider audience. Using deal.II's GPU features, we are planning an extension of the framework in this direction. As we saw in the results sections, the required resources that can be efficiently used vary over the layers: in the first layer, many processes do not receive any work and if they do, the communication overhead is too high to justify their use. Dynamic reallocation of more CPU cores as the problem domain grows would free the claimed but unused resources for other users of a compute cluster during the processing of the earliest layers.

	\appendix

	\section*{Acknowlegments}

	The authors would like to thank Katharina Kormann for her work on the matrix-free infrastructure and Marc Fehling for his work on the distributed hp-adpative infrastructure in \texttt{deal.II}. Furthermore, we thank Maximilian Bergbauer and Niklas Fehn for valuable discussions on high-performance computing and the \texttt{deal.II} matrix-free implementation.

	\section*{Author contributions}
	SP was responsible for the implementation of the model and the numerical studies. PM and MK supported the implementation and contributed general-purpose functionality to this project via the \texttt{deal.II} library. CM and WW are responsible for the conception of the modeling approach and acquisition of funding. All authors participated in writing and discussion of the manuscript.
	
	\section*{Funding}
	This work was supported by funding of the Deutsche Forschungsgemeinschaft (DFG, German Research Foundation) within project 437616465 and project 414180263.
	
	\section*{Supplementary material}
	
	\textbf{Video 1:} Scan-resolved thermal simulation of the manufacturing of all 312 layers of a cantilever specimen. The interlayer cool down phase of $1\,\si{\second}$ is not visualized.
	
	\newpage
	
	\appendix
	\section*{Appendix}
	\section{Temporal convergence}
	\label{sec:temporal_convergence}
\begin{figure}[h]
	 	\centering
	 	\includegraphics[width=.3\linewidth]{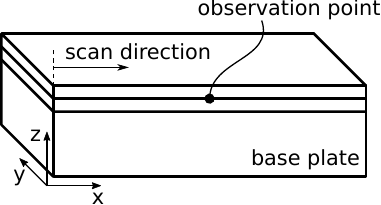}
	 	\caption{Example used to judge temporal convergence.}
	 	\label{fig:temporal_convergence_cube}
	 \end{figure}

\begin{figure}[h]
		\centering
		\includegraphics[width=\linewidth]{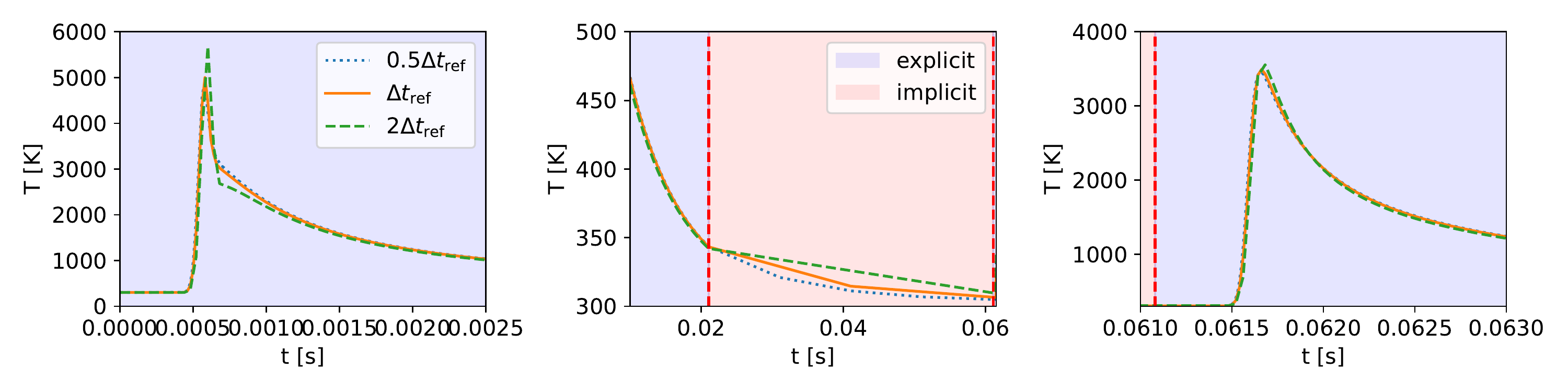}
       	\caption{Temporal convergence of temperature for the combined explicit-implicit time stepping scheme with different time step sizes. Left: detailed view of heat source moving over the observation point. Middle: switch from explicit to implicit scheme during cool down. Right: heat source in the second layer passes above the observation point.}
       	\label{fig:temporal_convergence}
	\end{figure}

	 A representative example of two layers with a single track each is used to analyze temporal convergence of the time stepping scheme for the parameters used in the numerical examples. The geometry consists of a $1.0\times 0.2 \times 0.2\, \si{\cubic\milli\meter}$ base plate ($x\times y\times z$ dimensions) with two powder layers of $40\, \si{\micro\meter}$ on top as illustrated in Figure~\ref{fig:temporal_convergence_cube}. Starting at ($x=0$, $y=0$), a single track is scanned along the $(1,0,0)$ direction in both layers with a $0.06\,\si{\second}$ cool down time in between layers. All other material and scan parameters are identical to Table~\ref{tab:mat_params} and Table~\ref{tab:scan_params_cantilever}. The temperature at the observation point $(0.5\, \si{\milli\meter},0.0, 0.24\, \si{\milli\meter})$ (top of first layer, middle of track) is shown in Figure~\ref{fig:temporal_convergence} for three different time step sizes, where $\Delta t_\text{ref}$ is the time step size used in the numerical examples ($\num{2e-5}\,\si{\second}$ in the explicit phase, $\num{2e-2}\,\si{\second}$ in the implicit phase). The switch between explicit and implicit time stepping and activation of a new layer is robust and the results are well converged for the desired level of accuracy in a part-scale model.
	 
    \color{black}

	\bibliographystyle{abbrv}
	\bibliography{ref}
\end{document}